 \documentclass[10pt]{iopart} 
\usepackage{iopams}  
\usepackage{graphicx}
\usepackage{siunitx} 
\usepackage{caption} 
\usepackage{subcaption} 

\begin{document}

\title[Comparison of Methods to Determine the Fluence of Monoenergetic Neutrons]{Comparison of 
Methods to Determine the Fluence of Monoenergetic Neutrons in the Energy Range from 30\,keV to 14.8\,MeV }

\author{R. Nolte$^1$\footnote{retired} and B. Lutz$^1$}
\address{$^1$ Physikalisch-Technische Bundesanstalt, Bundesallee 100, 38116 Braunschweig, Germany}
\ead{benjamin.lutz@ptb.de}

\begin{abstract}

The primary reference instruments for neutron fluence measurements used at the Physikalisch-Technische Bundesanstalt (PTB) 
are based on the primary standard for neutron measurements which is the differential 
neutron-proton scattering cross section. Such instruments require considerable effort for their
operation and analysis. Therefore, routine measurements are carried out using a transfer instrument 
to facilitate the efficient provision of services to customers. A series of measurements was conducted
to compare the transfer device to the primary reference instruments and ensure the traceability of neutron 
fluence measurements. This resulted in an improved characterization of the instrument and new analysis procedures.
For neutron energies between 144\,keV and 14.8\,MeV, the ratio of neutron fluence values measured with 
the primary reference instruments and the transfer instrument deviates from unity by less than the 
estimated standard measurement uncertainties of 2.6\,\% - 3.2\,\%. At neutron energies between 30\,keV and 100\,keV, however, 
the experimental fluence ratios deviate from unity by about 4\,\% which exceeds the estimated uncertainties of of 2.5\,\% - 2.9\,\%. At present, the reason 
for this inconsistency remains unresolved.

\end{abstract}


\submitto{\MET}
%
\ioptwocol

\section{\label{sec:Intro} Introduction}

At the PTB Ion Accelerator Facility (PIAF), neutron fields are generated for the characterization of 
neutron detectors and for various other applications. These range from measurements of nuclear data to the
irradiation of samples for radiobiological studies and radiation hardness testing. The basic quantity to 
be determined for these services is the neutron yield, that is, the number of neutrons 
$Y_\mathrm{n} = (\mathrm{d}N_\mathrm{n}/\mathrm{d}\Omega)(\Theta_\mathrm{n})$ emitted per unit solid angle 
from the production target at an angle $\Theta_\mathrm{n}$ relative to the direction of the ion beam. 
The yield is calculated from the fluence $\Phi_\mathrm{n}$ measured at a distance $d$ from the neutron production target:
$Y_\mathrm{n} = \exp(\Sigma_\mathrm{a}d)\,\Phi_\mathrm{n}\,d^2$. Here, $\Sigma_\mathrm{a}$ denotes the macroscopic total 
cross section of air and it is assumed that the fluence $\Phi_\mathrm{n}$ is already correctecd for inscattering of neutrons from air 
and structural materials.

Over a period of forty years, various methods have been implemented for  
measurements in monoenergetic and quasi-monoenergetic neutron fields at PIAF: the recoil proton 
proportional counter (RPPC) P2 \cite{Schlegel_2002-1}, the recoil proton telescope (RPT) T1\cite{Schlegel_2002-2} and the precision long 
counter (PLC) LC1 \cite{Hunt_1976}. In addition, \textsuperscript{235}U and \textsuperscript{238}U fission ionization chambers, 
several liquid scintillation detectors and \textsuperscript{6}Li glas scintillation detectors are in use at PIAF. 

The P2 RPPC and the T1 RPT are primary reference instruments  that allow the determination of the neutron fluence from
the mechanical dimensions of the instruments, the amount of hydrogen nuclei in the sensitive volumes and the 
differential neutron-proton (n-p) scattering cross section. This cross section is considered to be a primary standard for 
neutron metrology in the energy region from 1\,keV to 200\,MeV \cite{Carlson_2018}. 
For fast neutrons with energies above 100\,eV, elastic scattering is the only open channel in the neutron-proton system because the neutron 
capture cross section is more than four orders of magnitude smaller than the elastic scattering cross section. 
The n-p differential scattering cross section can 
be determined with a relative measurement of the angular distributions of the scattered neutrons 
and the recoil protons together with another relative measurement of the total hydrogen cross section.  
Therefore, ‘absolute’ neutron measurements can be carried out using these instruments. 
`Absolute’ here means that, in principle, only counting of events is needed. In reality, however, corrections are always required 
that compromise the ideal image.

In contrast to the primary reference instruments, the LC1 PLC is a transfer instrument which must be 
calibrated, for example using a \textsuperscript{252}Cf neutron source with known emission rate. In addition, the relative 
energy dependence of the fluence response must be calculated with the Monte Carlo method. The emission rate of \textsuperscript{252}Cf 
neutron sources can be measured using a manganese bath, which is a primary reference instrument as 
well. The PLC can therefore be used to link two primary methods of neutron measurement, namely neutron
fluence measurements relative to the n-p scattering cross section and the measurement of the emisssion of
radionuclide neutron sources using the manganese bath technique which is related to the radioactivity standards.
The uncertainty of the fluence response of the primary PTB reference instruments P2 and T1 is largely determined by the uncertainty 
of the n-p scattering cross section at the energy of the incident neutron. In contrast, the fluence response of 
the long counter is sensitive to the uncertainty of the hydrogen and carbon cross sections over the whole energy 
range from the energy of the incident neutron to thermal energies. A comparison with the primary reference instruments P2 and T1
can therefore be regarded as an integral experiment testing the overall uncertainty of the carbon and hydrogen 
cross sections.

The properties of the primary reference instruments impose restrictions on their practical use. This is especially true
in terms of the technical effort, the long measurement times required and the more complex
analysis, which makes the routine use of these instruments less favourable. In contrast, a PLC is a very 
stable and easy-to-use instrument for the time-efficient characterization of the neutron fields which 
helps users to make the best use of the allocated beam time. 
It is therefore important to demonstrate the traceability of LC1 measurements to the primary reference instruments.

An initial attempt to compare P2, T1 and LC1 was made during the last key 
comparison CCRI(III)-K11 \cite{Gressier_2014}. At this time, the PLC was employed in parallel to the RPPC and the RPT where possible, 
or it was used as a transfer instrument when the RPPC could not be employed for technical reasons \cite{Nolte_2018}.
The present work reports on a more extensive series of measurements aiming at a comparison of the LC1 transfer 
instrument and the primary reference instruments P2 and T1 in the neutron energy range from 30\,keV 
to 14.8\,MeV. This extends the measurements already performed within \mbox{CCRI(III)-K11} and 
provides data for crosschecking of the relative energy dependence of the PLC fluence response 
calculated with the Monte Carlo method. 

The analysis procedures reported here deviate from the `standard' procedure discussed in the PTB report \cite{Nolte_2018} for CCRI(III)-K11. 
This pertains to enhancing the characterization of instruments, calculating corrections, and utilizing state-of-the-art Monte Carlo codes 
to simulate the instruments.

In the present work, the differential n-p scattering cross section $\sigma_\mathrm{np}$
of the \mbox{ENDF/B-V} evaluation \cite{Uttley_1983} is still used for the analysis of the measurement with the primary
reference instruments. This follows an agreement within the CCRI(III) committee to not accept
the unmotivated increase of the differential \mbox{n-p} scattering cross section at backward angles in the energy range 
between 10\,MeV and 15\,MeV introduced in the \mbox{ENDF/B-VI} evaluation and later removed again 
to large extend in the \mbox{ENDF/B-VII} evaluation. 

\begin{figure}[htbp]
  \includegraphics[width=0.45\textwidth]{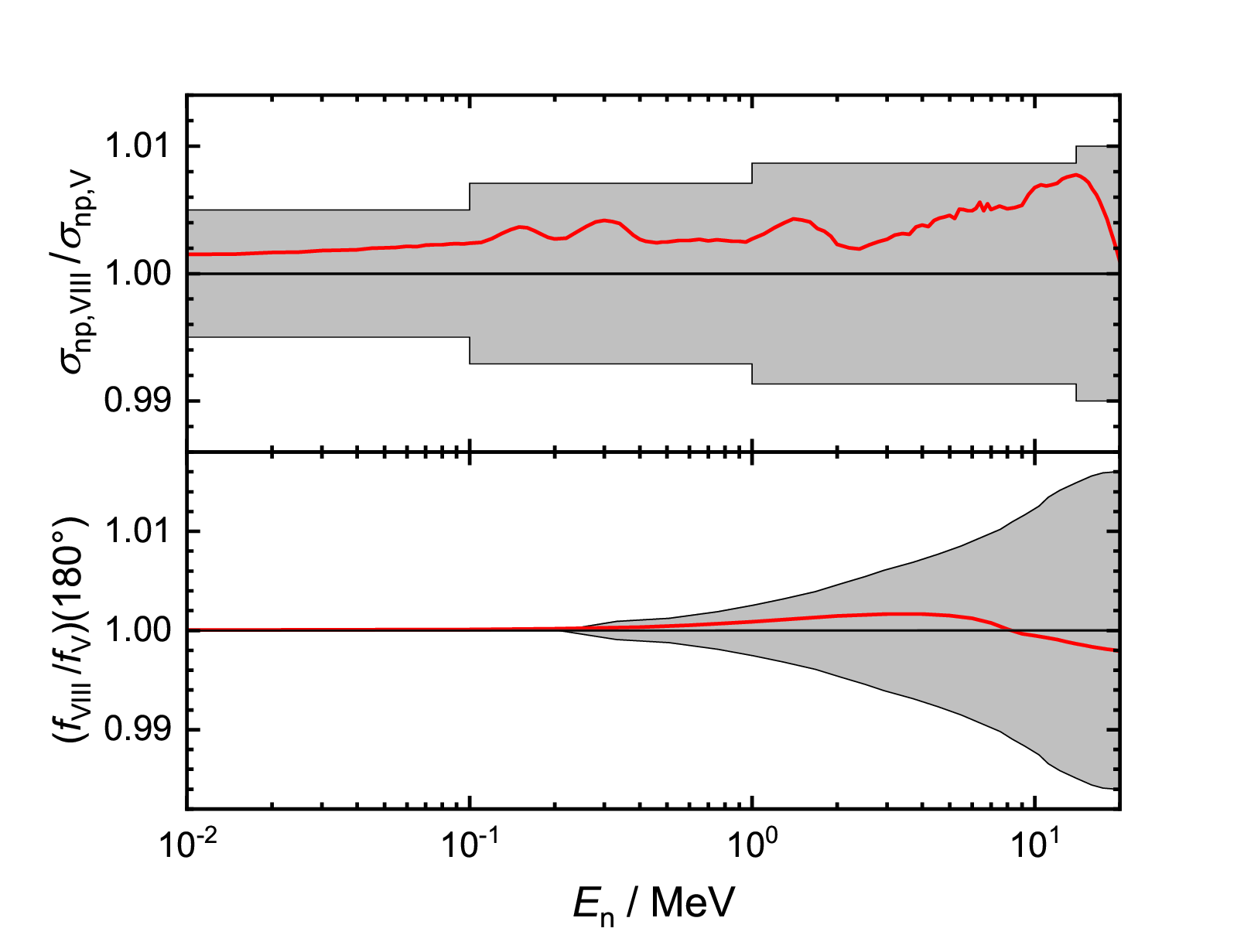}
  \centering
  \caption{\label{fig:XSratio} 
  The red solid lines show the ratios of the n-p scattering cross section $\sigma_\mathrm{np}$ (upper panel) and backward 
  angular distributions $f(\ang {180})$ (lower panel) from the \mbox{ENDF/B-VIII} and \mbox{ENDF/B-V} evaluations. The uncertainty 
  band (grey) in upper panel shows $1 \pm u_\mathrm{np,rel}$, where $u_\mathrm{np,rel}$ denotes the relative uncertainty of the \mbox{ENDF/B-V} 
  cross section used in the present study. 
  Due to a lack of covariance data for the \mbox{ENDF/B-V} differential cross section, the uncertainty band in the lower panel was obtained 
  from the relative uncertainty of the \mbox{ENDF/B-VIII} backward diffferential cross section \cite{Hale_2019}.}
\end{figure}

As shown in the upper panel of figure\,\ref{fig:XSratio},  
the deviation between n-p cross sections from the most recent evaluation \mbox{ENDF/B-VIII} \cite{Carlson_2018} and the
\mbox{ENDF/B-V} evaluation is smaller than the uncertainty of 
the \mbox{ENDF/B-V} data used in the present study. The lower panel compares the backward angular distribution 
$f(\ang{180}) = 2 \pi \sigma^{-1}_\mathrm{np} (\mathrm{d} \sigma_\mathrm{np} / \mathrm{d} \Omega) (\ang{180})$
of the differential scattering cross section in the center-of-mass system that is required for the analysis of the T1 measurements 
(see \ref{subsubsec:RPT-Analysis}). Unfortunately, an estimate of the uncertainty of the 
differential center-of-mass scattering cross section at \ang{180} is only available for the \mbox{ENDF/B-VIII} evaluation \cite{Hale_2019}. 
Therefore, the relative uncertainty of the backward angular distribution $f(\ang{180})$ was obtained by quadratically 
subtracting the relative uncertainty of the \mbox{ENDF/B-VIII} scattering cross section $\sigma_\mathrm{np}$ from that of the 
\mbox{ENDF/B-VIII} backward differential cross section $(\mathrm{d} \sigma_\mathrm{np} / \mathrm{d} \Omega)(\ang{180})$.
This is a lower bound for the uncertainty of the \mbox{ENDF/B-V} angular distribution at \ang{180}. 
Also here, the deviation between the two evaluations is smaller than the estimated uncertainty of the backward scattering cross section.

This study reports the results for the energy region between 30\,keV and 15\,MeV \footnote{All uncertainties indicated in this publication
are standard measurement uncertainties with a coverage factor $k = 1$.}.
The energy range from 
15\,MeV to 19\,MeV will be covered in a forthcoming publication. The monoenergetic neutron fields 
are described in section\,\ref{sec:Fields}. Section\,\ref{sec:Methods} 
gives a summary of the properties of the instruments and the analysis methods. The 
results of the comparison are presented and discussed in the final sections.

\section{\label{sec:Fields} Monoenergetic Neutron Fields}

\subsection{\label{subsec:Production} Neutron Production}

The measurements reported here were carried out in the PIAF low-scatter hall during 
two periods from 2014 to 2016 and 2020 to 2023. Monoenergetic neutrons were generated using 
proton and deuteron beams from the 4\,MV Van-de-Graaff accelerator (2014 – 2016), the 2\,MV 
Tandetron accelerator (2020 – 2023) and the CV28 cyclotron together with the nuclear reactions  
\textsuperscript{7}Li(p,n)\textsuperscript{7}Be (30\,keV - 750\,keV), \textsuperscript{3}H(p,n)\textsuperscript{3}He 
(750\,keV - 4\,MeV), D(d,n)\textsuperscript{3}He (4\,MeV - 8\,MeV)
and \textsuperscript{3}H(d,n)\textsuperscript{4}He (14.8\,MeV - 20\,MeV). 

The proton and deuteron energies of the beams produced with the Van-de-Graaff and the Tandetron accelerators were determined with a \ang{90} 
analysing magnet that was calibrated using a set of (p,$\gamma$) 
resonances and (p,n) reaction thresholds. Nevertheless, deviations between the preset and actual beam energy of 
2\,keV to 6\,keV were observed frequently during routine operation of the accelerators and the beam 
transport system. 
These deviations would lead to an unacceptable uncertainty in the mean 
neutron energy when the beam energy is close to the reaction threshold. Consequently, for neutrons with energies 
between 144\,keV and 500\,keV produced by the \textsuperscript{7}Li(p,n)\textsuperscript{7}Be reaction, 
the mean neutron energy $\bar{E}_\mathrm{n}$ of the monoenergetic peak was determined with a cylindrical \textsuperscript{3}He 
proportional counter (outer diameter: 50\,mm, outer length: 346\,mm, \textsuperscript{3}He pressure: 0.1\,MPa).
The neutron energy is obtained from the $Q$-value of the \textsuperscript{3}He(n,p)\textsuperscript{3}H reaction and the 
location of  two peaks in the pulse-height 
distribution originating from thermalized room-return neutrons and from the direct neutrons.
This method yields the mean neutron energy with a relative uncertainty of about 1\,\%, that is, 1.4\,keV 
for a mean neutron energy of 144\,keV.
The energies of the deuteron beams from the cyclotron were analysed using two \ang{90} dipole magnets. 
The deuteron energy spread was about 20\,keV for these beams.

\subsection{\label{subsec:Targets} Targets}

The lithium targets were 
prepared at the  PIAF target laboratory by physical vapour deposition (PVD) of \textsuperscript{nat}Li layers 
on 0.5\,mm tantalum backings. To prevent the lithium surface from corroding, the lithium targets were handled
without exposing the target to air from production to mounting on the beam line.
Neutron fields with energies above 100\,keV were always 
produced at a neutron emission angle of \ang{0} with respect to the ion beam direction. For
energies between 100\,keV and 30\,keV, the \textsuperscript{7}Li(p,n)\textsuperscript{7}Be reaction was used at neutron 
emission angles up to \ang{72}. In these cases, the mean zero-degree neutron energy was always 144\,keV.
By using lithium targets instead of the more readily available lithium fluoride targets, the intense 
production of parasitic high-energy photons by the \textsuperscript{19}F(p,$\alpha\gamma$)\textsuperscript{16}O 
reaction is avoided. This helps to reduce the interference of photon and neutron-induced events in the RPPC.  

The tritium targets 
consisted of tritium gettered in titanium layers that were produced by PVD on 0.5\,mm silver or 
aluminum backings. These targets were obtained from a commercial supplier and stored in vacuum when not in use.

To spread out the heat deposited in the target by the ion beam, the solid-state targets perform an oscillatory movement 
perpendicular to the ion beam at a frequency of about 1\,Hz. Due to the curved target surface, this movement 
induces a variation of the distance from a reference point to the beam spot on the reactive target layer. 
The distribution of the distances has a standard deviation of less than 0.3\,mm. This is of relevance for the use of the 
RPT T1 at small distances from the target (see \ref{subsubsec:RPT-Design}).

Neutron fields with energies between 5\,MeV and 8\,MeV were generated using a D\textsubscript{2} gas target.
This deuterium gas target is a 30\,mm long stainless-steel cell with a diameter of 11\,mm and a wall thickness of 0.2\,mm. 
The entrance window is covered by a 5\,\unit{\micro m} molybdenum foil which can stand gas pressures up to 0.2\,MPa.
The deuteron beam is stopped in a 0.5\,mm thick gold foil at the rear end of the cell.

\subsection{\label{subsec:Scattering} Target Scattering}

The quantity to be measured in a neutron field is the fluence $\Phi_\mathrm{dir}$ of neutrons leaving the target without 
further interaction (direct neutrons) \cite{Nolte_2011}. Therefore, the contribution of neutrons interacting with the target setup 
must be included in the calculation of the response of the reference instruments and in the analysis of measurements.
It should be noted here that this includes also that part of the energy distribution $(\mathrm{d}\Phi_\mathrm{sc}/\mathrm{d}E_\mathrm{n})$ of 
the scattered neutrons which falls into the energy region covered by the direct neutrons.

At PIAF, a low-mass setup is used for the solid-state targets which restricts the relative number of neutrons scattered 
in the target setup (scattered neutrons) to less than 10\,\%. Because of the high-density backings, the most problematic 
targets in this respect are the lithium targets on 0.5\,mm tantalum backings. The energy distribution of the direct neutrons 
is determined from a simulation of the ion beam transport and the neutron production in the target layer using the evaluated differential neutron 
production cross sections. This is followed by the simulation of neutron scattering in the target setup.

\begin{figure}[htbp]
  \centering
  
  \begin{subfigure}[b]{0.45\textwidth}
    \includegraphics[width=\textwidth]{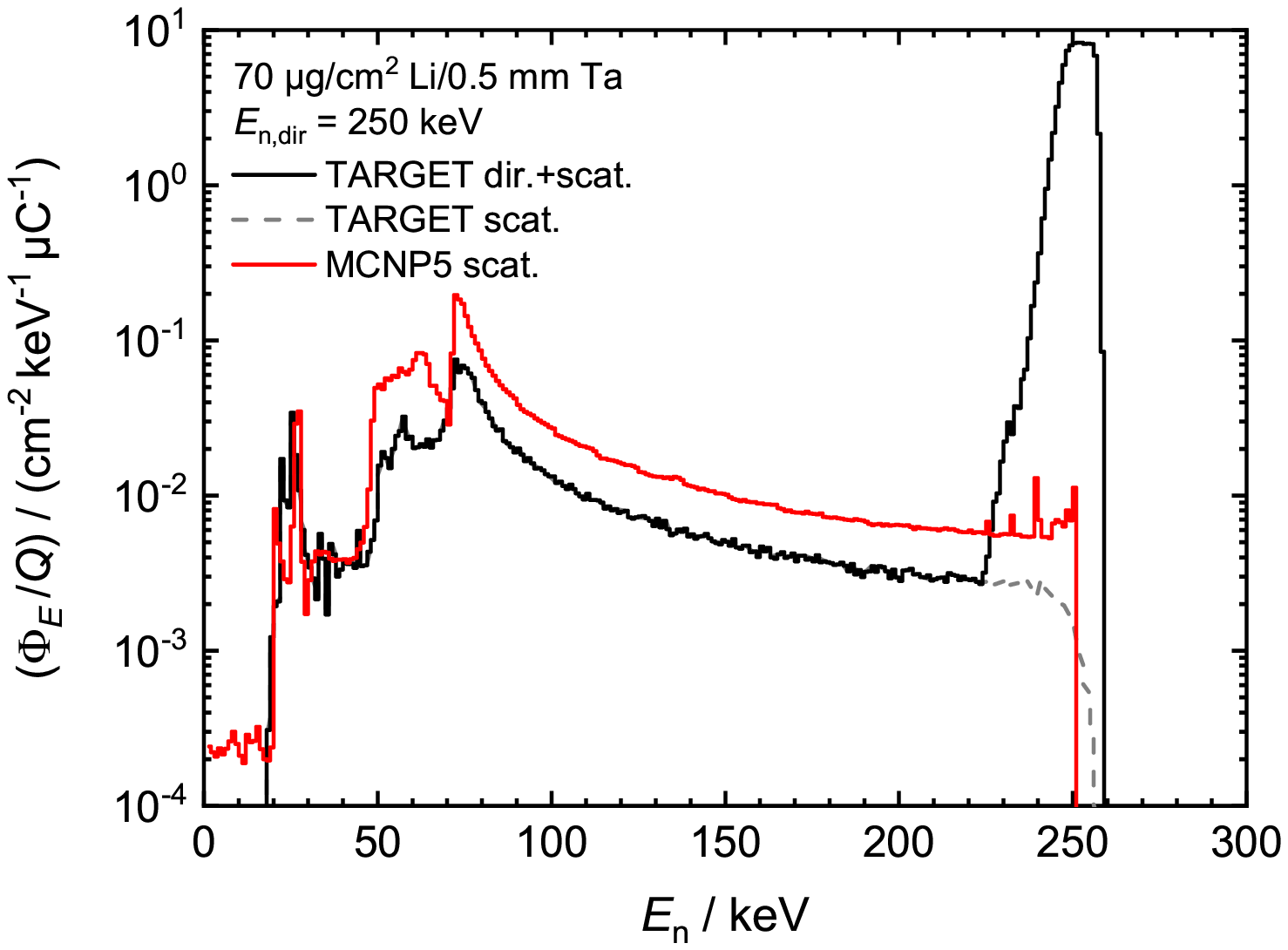}
  \end{subfigure}

  \begin{subfigure}[b]{0.45\textwidth}
    \includegraphics[width=\textwidth]{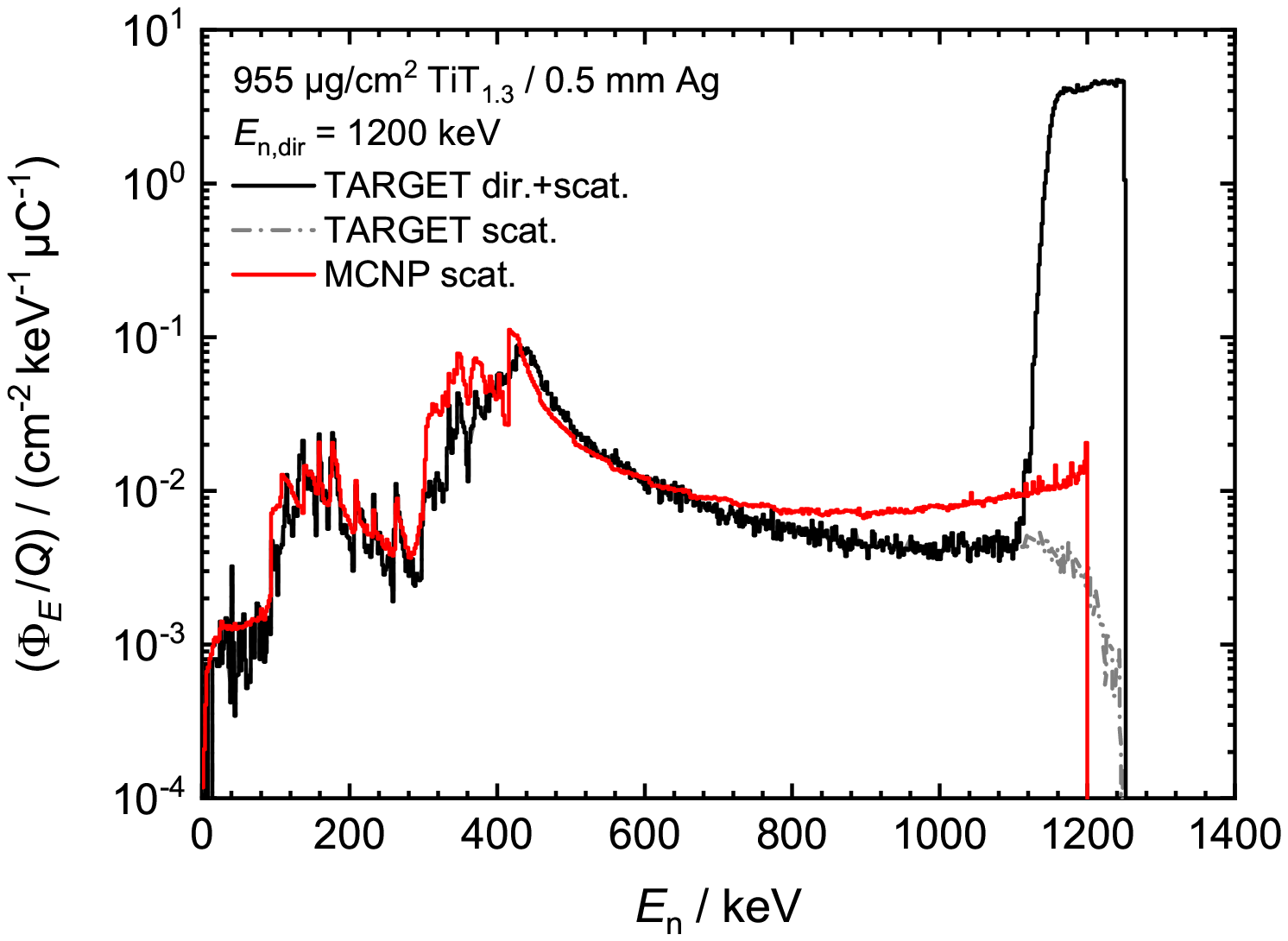}
  \end{subfigure}

  \centering
  \caption{\label{fig:Ti(T)andLitargets} 
  Energy distributions for the 250\,keV (upper panel) and 1.2\,MeV (lower panel) neutron fields calculated 
  using TARGET (black: direct and scattered neutrons, grey: scattered neutrons only) and MCNP5 (red: scattered neutrons only). 
  For the 250\,keV field, 
  the ratios of the scattered and the direct neutron fluence is 2.6\,\% (TARGET) and 5.4\,\% (MCNP5). For the 1.2\,MeV 
  field, the ratios is 3.4\,\% (TARGET) and 4.3\,\% (MCNP5).}
\end{figure}

The TARGET Monte Carlo code \cite{Schlegel_2005} provides the energy distributions of direct and scattered neutrons 
in a very time-efficient way by using point detectors,
but it has several known shortcomings. Therefore, the energy distributions of scattered neutrons were 
crosschecked using MCNP5 \cite{X-5_2004} with \mbox{ENDF/B-VII} cross section data. Figure\,\ref{fig:Ti(T)andLitargets} shows two 
typical cases. While the relative number of scattered neutrons predicted by TARGET is about 20\,\% below that from 
MCNP5 for the fields produced with tritiated titanium targets, deviations between 50\,\% and 60\,\% were observed 
for the fields generated with the lithium targets. 
It is most probable that the discrepancy is due to deficient differential cross section data for elastic scattering 
from tantalum. As discussed in \ref{subsec:RPPC}, this procedure is supported by results of measurements 
using the RPPC. 

For lithium targets, TARGET was therefore only used to calculate the energy distribution 
of the direct neutrons. The energy distribution of 
scattered neutrons was obtained with the MCNP5 Monte Carlo code, using the differential neutron production cross section data from
the DROSG2000 database \cite{Drosg_2016} to model the source term. This source term had a unique 
relation between neutron energy
and the emission angle, meaning that the energy loss of the ions in the reactive layer was neglected. A comparison 
with more elaborate source descriptions \cite{Birgersson_2009} showed that this was fully sufficient for the energy 
distribution of the scattered neutrons. The two calculations were normalized to the same yield of 
direct neutrons at zero degree. 

For the tritiated titanium targets on silver and aluminum backings, 
TARGET was used to calculated both distributions. The uncertainty of this procedure was estimated 
to be about 50\,\% of the difference between TARGET and MCNP5.

\begin{figure}[htbp]
  \includegraphics[width=0.45\textwidth]{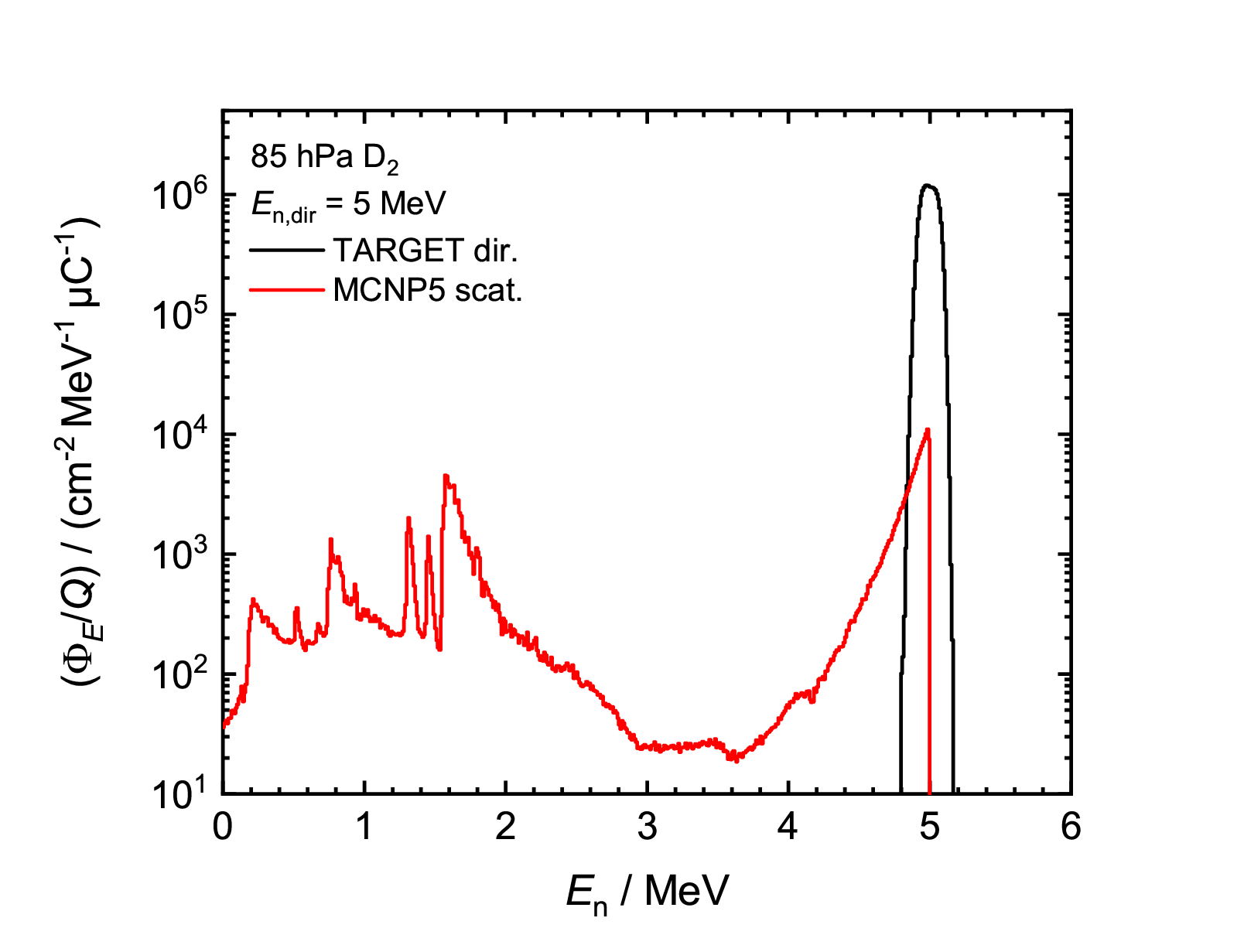}
  \centering
  \caption{\label{fig:D2target} 
  Energy distributions for the 5\,MeV neutron field target calculated using 
  TARGET (black histogram) and MCNP5 (red histogram, scattered neutrons only). The ratios of the scattered and 
  the direct neutron fluence calculated with MCNP5 is 1.5\,\%. The TARGET code does not provide the energy 
  distribution of scattered neutrons. The neutron field was produced with the D\textsubscript{2} gas  target.}
\end{figure}

Due to the forward-peaked differential cross section of the D(d,n)\textsuperscript{3}He reaction and the geometry of the gas target, 
the relative contribution of scattered neutrons in the neutron fields produced with the D\textsubscript{2} gas target is significantly less 
than for the solid-state targets. Consequently, the TARGET code does not provide the energy distribution of the
scattered neutrons for these fields. For the present study, the relative contribution of the scattered neutrons has been calculated 
in the same way as for the lithium targets. It varied between 1.5\,\% for the 5 MeV field  and 1.1\,\% for the 8 MeV field. 
Figure\,\ref{fig:D2target} shows the neutron energy distribution for the 5\,MeV neutron field.

The energy distributions of the direct neutrons calculated with the TARGET code are averaged over the sensitive cross-sectional
area of the detector. The corresponding opening angles covered by the instruments at the closest distances to the target were
\ang{1.4} (P2 at 115.5\,cm), \ang{3.5} (T1 at 22.6\,cm) and \ang{4.4} (LC1 at 247.3\,cm). This approximation introduces 
deviations of less than 0.3\,\% from the result obtained when the angular distribution within the solid angle 
covered by the instruments is included in the simulation.
For the calculation of the response of P2, T1 and LC1 using the Monte Carlo codes MCNP5 \cite{X-5_2004} and MCNPX \cite{MCNPX_2011}, the angular
distributions of scattered and direct neutrons with the \textsuperscript{7}Li(p,n)\textsuperscript{7}Be, 
T(p,n)\textsuperscript{3}He and T(d,n)\textsuperscript{4}He reactions were therefore assumed to be isotropic 
over the range of neutron emission angles covered by the instruments. 
Only for the D(d,n)\textsuperscript{3}He reaction the angular distribution of the 
direct neutrons was modelled using the differential cross section at the deuteron energy corresponding to the mean energy
of the direct neutrons at \ang{0}. This was necessary because the angular distribution of this reaction is much steeper 
than those of the others.

In addition to target scattering, neutrons are also scattered from air and other structural materials before being detected in 
the reference instruments. The contribution of these \lq{}room-return\rq{} 
neutrons to the measurements with the P2 RPPC and the LC1 PLC were subtracted experimentally using the shadow cone technique. 
Due to the short distance between the T1 RPT and the target, the effect of room-return neutrons was negligible for this 
instrument. Outscattering of neutrons between the production target an the sensitive volume was included in the 
Monte Carlo simulations by including a cone filled with air which had the same opening angle and position of 
the apex as the physical shadow cone. 

\subsection{\label{subsec:Monitors} Monitors}

\begin{figure}[htbp]
  \centering
  \begin{subfigure}[b]{0.45\textwidth}
    \includegraphics[width=\textwidth]{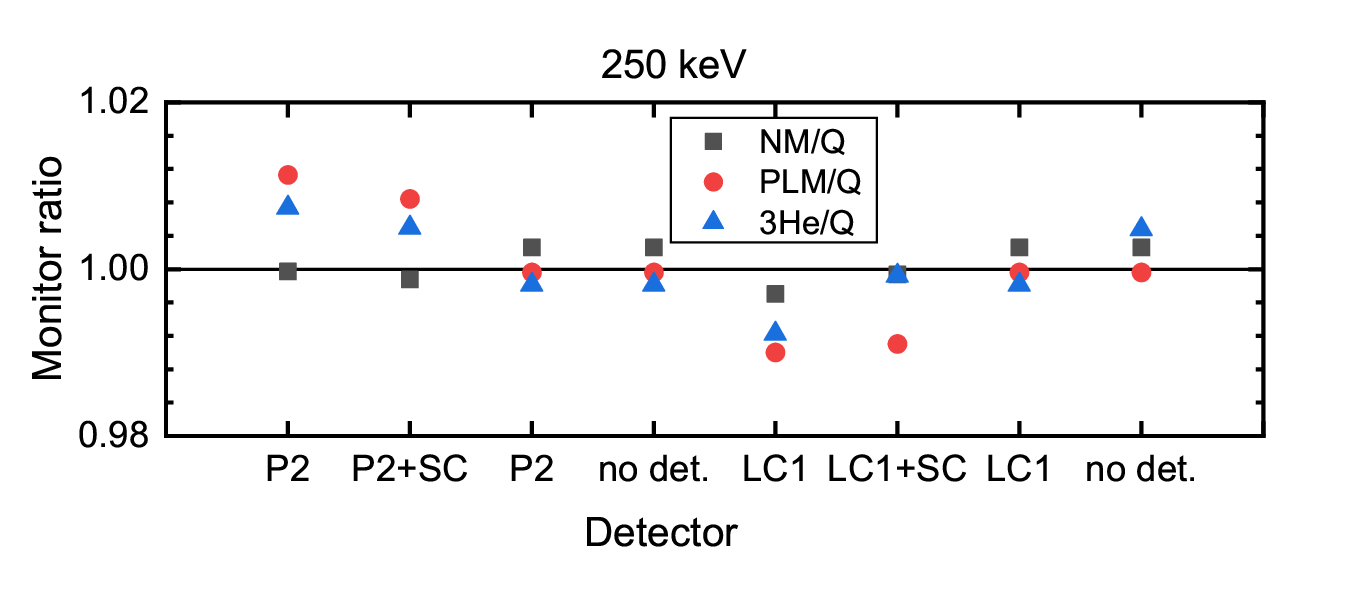}
  \end{subfigure}

  \begin{subfigure}[b]{0.45\textwidth}
    \includegraphics[width=\textwidth]{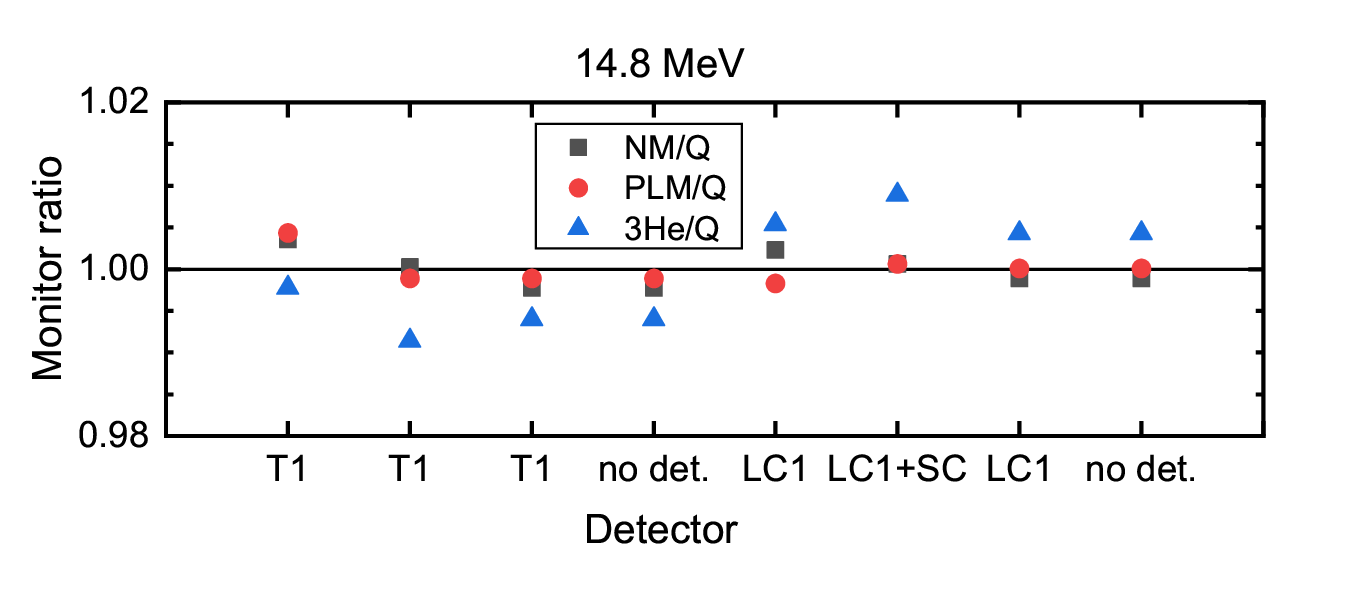}
  \end{subfigure}

  \caption{\label{fig:Monitorratios} 
  Normalized ratios of the number of neutron monitor events to the integrated beam current $Q$. 
  For the measurements marked with `no det.', all reference instruments were removed from the neutron field to 
  suppress scattering from these instruments into the monitor detectors. 
  Measurements with the shadow cone in place are denoted by `+SC'.  The relative standard deviations
  of the NM/Q ratios for the data sets at 250\,keV and 14.8\,MeV were 0.22\,\% and 0.18\,\%, respectively}
\end{figure}

Measurements with different detectors were related to each other by simultaneously recording the integrated beam current $Q$ and the number of 
events from a set of three neutron monitors: a \textsuperscript{3}He long counter (NM) at \ang{19}, a BF\textsubscript{3} 
long counter (PLM) at \ang{97} and a \textsuperscript{3}He counter (3He) in a cover of borated polyethylene at 
an angle of about \ang{140}. The distances of the neutron monitors NM, PLM and 3He 
from the target were 6.05\,m, 5.30\,m and 0.80\,m, respectively. 
For most beam settings, the relative standard 
deviation of the ratio of the number of neutron monitor events to the integrated beam current was 
less than 0.5\,\% during one measurement period with the same ion beam and neutron 
production target. For the majority of the measurements, the NM monitor was the most stable one. Therefore this monitor was 
usually used to analyse the measurements. The scattering of neutrons from other detectors 
into the monitors was corrected for by performing short 
subsequent runs with the respective detector at its measurement position and with the detector removed from the neutron field. 
During these runs, the integrated beam current $Q$ was used as a monitor to determine the number of additional 
monitor events induced by neutrons scattered from the detector. For the detectors employed in this study, the 
relative effect of inscattering ranged between 1\,\% and 2\,\% at maximum and was rather independent of 
the neutron energy. Figure\,\ref{fig:Monitorratios} shows two sets of normalized monitor ratios for 
measurements at 250\,keV and 14.8\,MeV.

\section{\label{sec:Methods} Methods}

\subsection{\label{subsec:RPPC}Recoil Proton Proportional Counter P2}

\subsubsection{\label{subsubsec:RPPC-Design}Design of the Instrument}

\begin{figure}[ht]
  \includegraphics[width=0.45\textwidth]{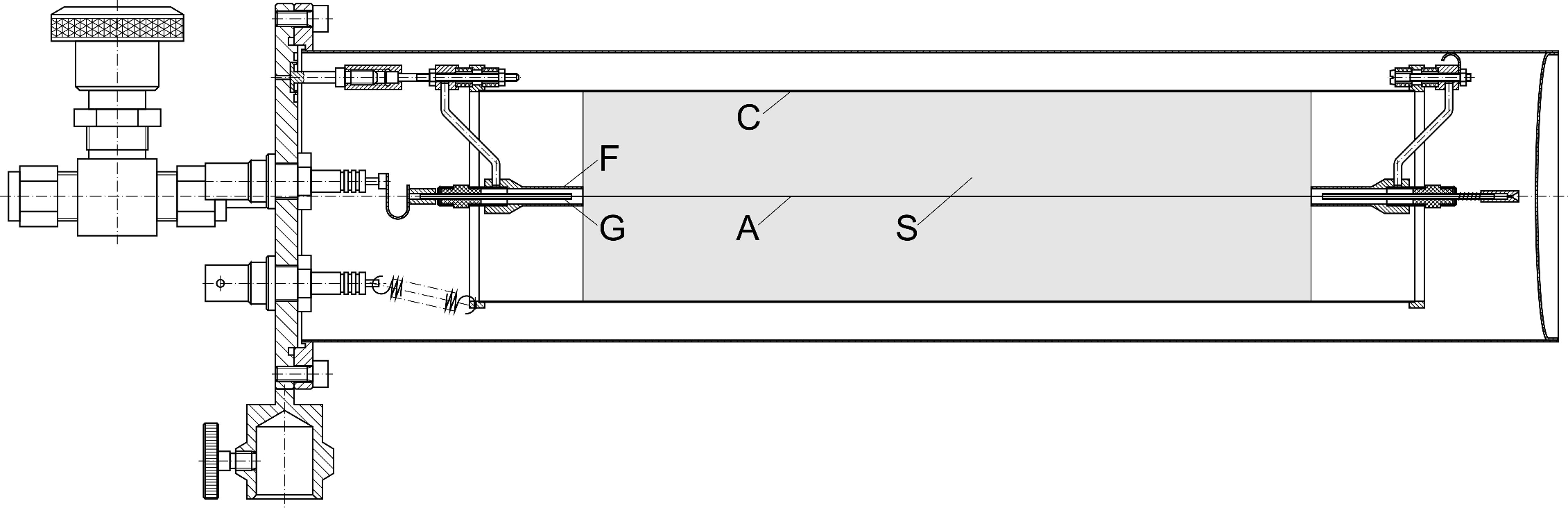}
  \centering
  \caption{\label{fig:P2}
  The P2 recoil proton proportional counter is used as the primary reference instrument for the neutron energy range 
  from 24\,keV to 2\,MeV. The geometrical sensitive volume S is shaded in grey; the cathode is denoted by C 
  and the anode wire by A.  Field and guard tubes are marked by F and G, respectively. The guard tubes are held at 
  anode potential while the field tubes are at ground potential. The ratio of anode and cathode potentials is 1.402. 
  The anode wire is a gold-plated tungsten wire 100\,\unit{\micro m} in diameter.}
\end{figure}

The P2 RPPC has been in use at PTB since the 1980s. As shown in figure\,\ref{fig:P2}, its design closely follows the 
original layout of Skyrme \etal \cite{Skyrme_1952}. The instrument is employed in the energy region 
between 24\,keV and about 2\,MeV. 
It consists of a 0.5\,mm thick cylindrical stainless-steel housing, 76\,mm in diameter 
and 360\,mm in length. The thickness of the stainless-steel entrance window is 0.5\,mm as well. The 
mechanical size of the sensitive volume within this housing is defined by a cylindrical aluminum 
cathode, 55.52(5)\,mm in diameter and 0.3 mm in thickness. The height of the volume is 193.3(2)\,mm and it is 
defined by the end of the field tubes surrounding the anode wire with a diameter of 100 µm. 
The field tubes are held at ground potential while the anode and cathode potentials are kept at a 
voltage ratio $U_+ / U_-$ of  1.402, as calculated from the diameters of the anode wire, field tubes and cathode.   

The counter is positioned with the cylindrical axis on the zero-degree direction of 
the neutron field and at a distance of about 115.5\,cm between the neutron production target and 
the geometrical centre of the sensitive volume.  The contribution of neutrons scattered by air or 
by structural materials of the low-scatter facility was subtracted experimentally by conducting
separate measurements with a 30\,cm long polyethylene shadow cone placed between the RPPC and the 
neutron production target.

\subsubsection{\label{subsubsec:E-Field}Electrical Field}

\begin{figure}[ht]
  \includegraphics[width=0.35\textwidth]{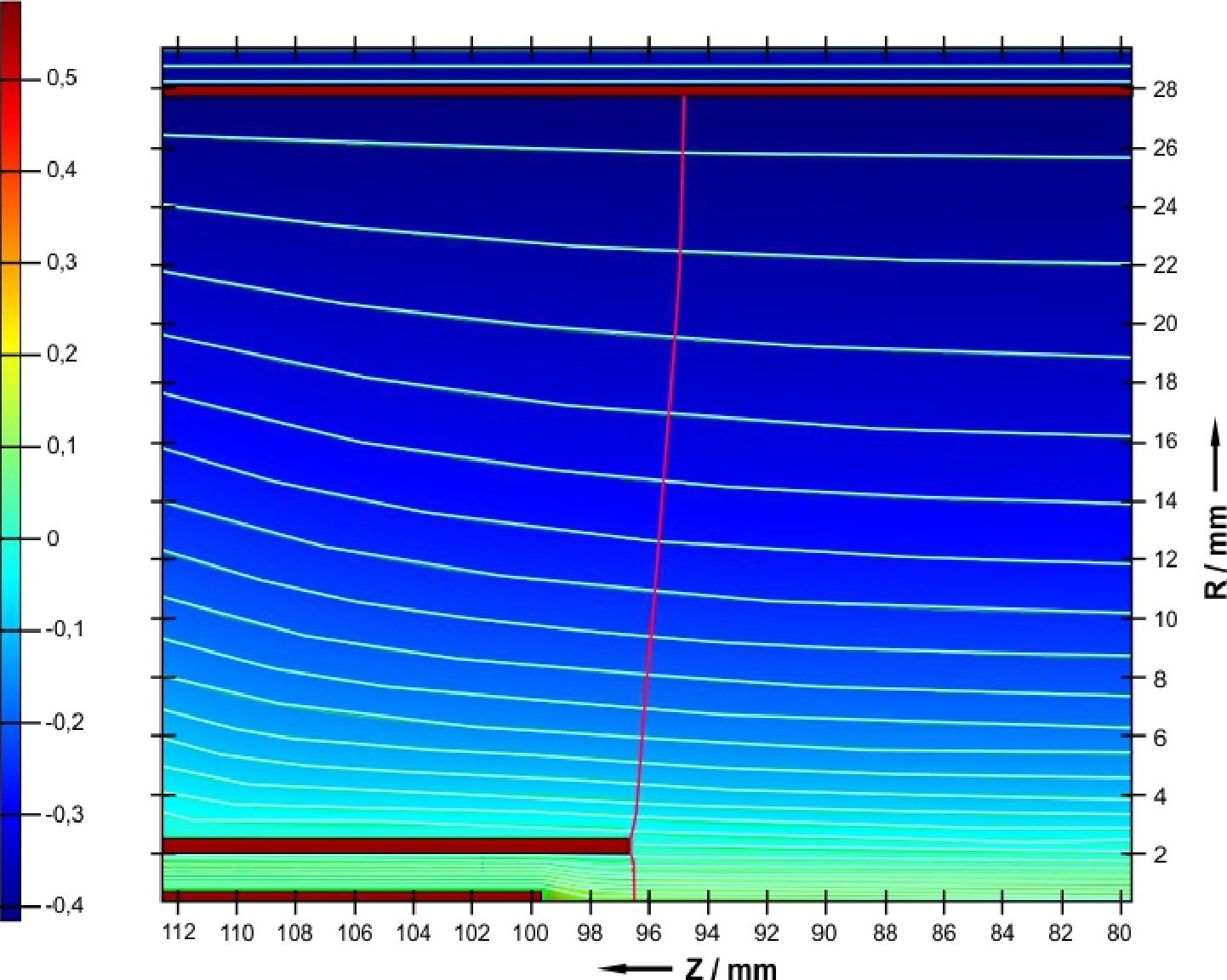}
  \centering
  \caption{\label{fig:E-field}
  Equipotential surfaces (light blue lines) inside P2 and the electrical field line (red line) 
  touching the end of the field tube. This field line defines the 
  surface of the sensitive volume. The axial and radial coordinates $Z$ and $R$ are the distances from the 
  center of the sensitive volume and the counting wire, respectively. 
  The electrical field was calculated with COMSOL Multiphysics \cite{COMSOL_2014}.}
\end{figure}

The efficiency of the instrument is defined by the size of the volume from which electrons can drift to the amplification region around the 
anode wire. When neglecting lateral diffusion during the drift of the electrons along the field lines, 
this volume is defined by the electric field at the axial heights of the ends of the field tubes. 
The electric field distribution was calculated numerically using COMSOL 
Multiphysics \cite{COMSOL_2014}. Figure \,\ref{fig:E-field} shows a field map. In the vicinity of the end of the field tubes the 
equipotential surfaces are not parallel to the anode wire. Therefore, the separatrix is not a plane surface perpendicular to the axis of the 
geometrical volume. A correction factor $k_\mathrm{V} = 0.985(8)$ was therefore applied to the geometrical volume 
$V_\mathrm{geo} = 4.680(10)\,\times 10^2\,\mathrm{cm}^3$. 

\subsubsection{\label{subsubsec:Gas}Counting Gas}

The gas types and pressures used in the RPPC are selected such that the leakage of recoil protons 
from the sensitive volume is kept within acceptable limits. For neutron energies below 400 keV, a 
mixture of 96.5\,\% hydrogen and 3.5\,\% methane per unit volume was used, and propane was employed at energies 
above 400\,keV. The pressure $p$ varied between 300\,hPa and 1000\,hPa, depending on the gas type 
and neutron energy. 

\begin{figure}[ht]
  \includegraphics[width=0.45\textwidth]{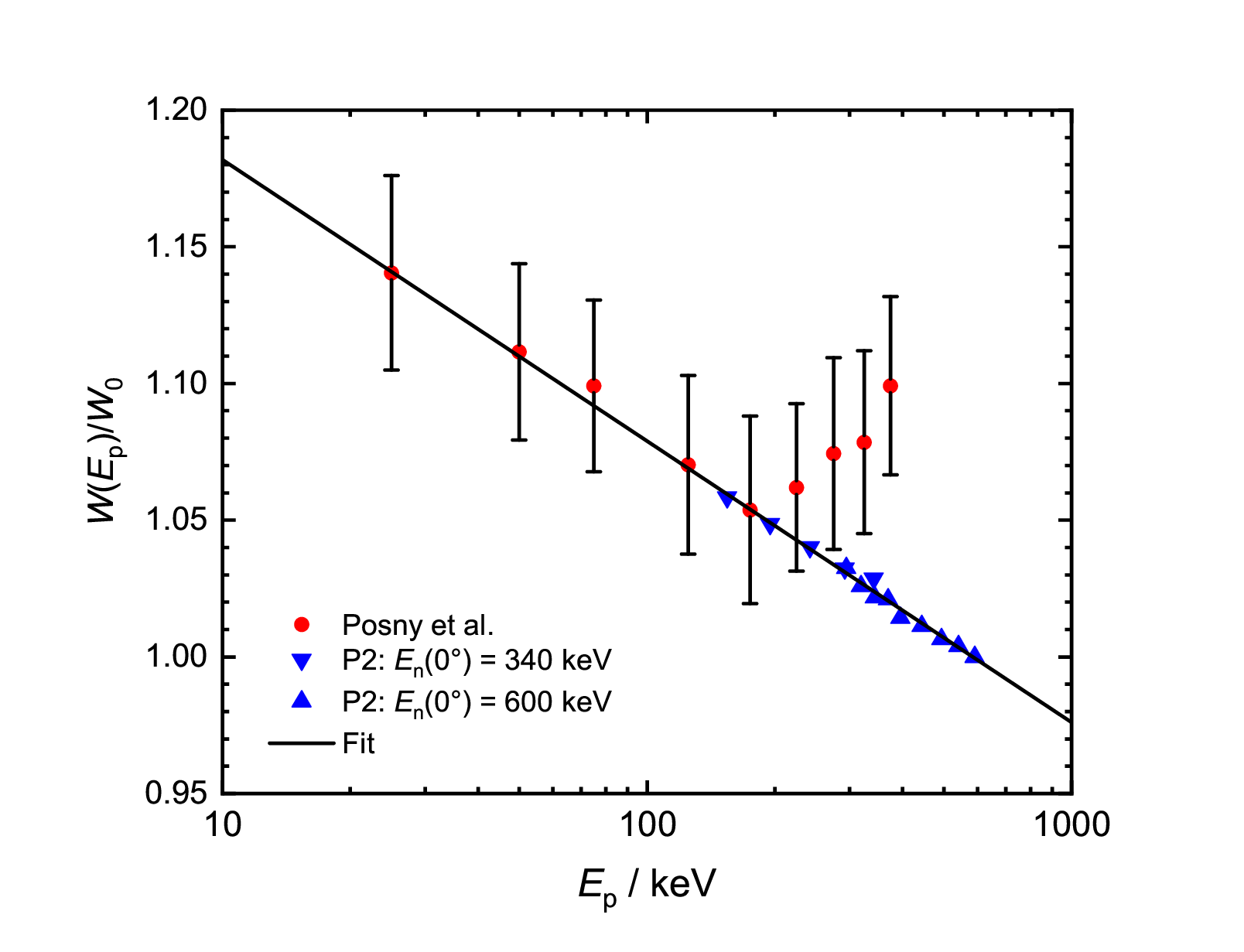}
  \centering
  \caption{\label{fig:W-valuedata}
  Energy dependence of the relative \textit{W}-value for protons in propane. The red circles show measurements of 
  the absolute \textit{W}-value by Posny \etal \cite{Posny_1987}.
  The data shown by the blue triangles are relative \textit{W}-values normalized to the Posny data at 200\,keV. 
  They were obtained with the P2 RPPC by measuring the variation of the recoil proton edge with neutron energy. 
  The neutron energy was varied by changing the angular position of the RPPC.
  The linear fit on the logarithmic energy scale (black solid line) was used 
  for analysing the PTRAC file produced by MCNPX.}
\end{figure}

The conversion of proton energy deposited in the 
gas to ionization charge is described by the $W$-value.  
For propane an energy dependence $W(E_\mathrm{p})$ was considered, where $E_\mathrm{p}$ denotes the kinetic 
energy of the recoil protons. This was based on the data of Posny \etal 
\cite{Posny_1987} and on two measurements of the relative $W$-value $W(E_\mathrm{p})/W_0$ 
with the P2 RPPC, where $W_0$ denotes the $W$-value for 600\,keV protons. These measurements 
were made by varying the neutron energy and observing the change of the location of the recoil proton edge 
in the pulse height distribution. For the simulations, the data  in figure \ref{fig:W-valuedata} 
were approximated by a linear relation on the logarithmic energy scale. This fit excluded the 
Posny data above 200\,keV because the re-increase of these data could not be verified in the 
measurements with P2. In contrast to propane, there is no evidence of a significant energy dependence of 
the $W$-value for pure hydrogen and  hydrogen-methane mixtures for protons energies above 10\,keV 
\cite{Breitung_1972}, \cite{Grosswendt_1997}, that is, $W(E_\mathrm{p})/W_0 = 1$ for the hydrogen-methane mixtures. 

\subsubsection{\label{subsubsec:RPPC-Simulation}Monte Carlo Simulation}

The neutron response of the counter was calculated using MCNPX, v2.7 \cite{MCNPX_2011}. Cross section 
data from the ENDF/B-VII evaluation were used except for the n-p cross section that was taken from \mbox{ENDF/B-V}. 
The simulated neutron energy distributions discussed in section\,\ref{sec:Fields} were used to model the source term. 
The simulations were carried out for a divergent source at a distance $d_0$ of 115.25 cm 
from the geometrical center of the RPPC and with a layer of air between the source and the RPPC to 
account for outscattering of neutrons in air. The angular distribution of the source was assumed to be 
isotropic because the opening 
angle covered by the sensitive volume is only \ang{1.5}. The PTRAC files produced by MCNPX were analyzed 
using a dedicated filter code to determine the `pulse height' 
$(W(E_\mathrm{p})/W_0) E_\mathrm{p}$ produced by recoil protons in the sensitive 
volume. In this way all neutron and proton transport effects were correctly modelled. 
The finite pulse-height resolution of the RPPC was accounted for by folding the 
simulated pulse height distributions with Gaussians of constant relative widths.
Unfortunately, recoil nuclei with $Z > 2$ are not transported by MCNPX. Therefore, the contribution from 
carbon recoil nuclei at a low pulse height could not be modelled and had to be excluded from the analysis.  
The simulated pulse height distributions were normalized to the unit fluence of direct neutrons 
$\Phi_\mathrm{dir}$ in vacuum at the position of the center of the sensitive volume.

\begin{figure}[ht]
  \includegraphics[width=0.45\textwidth]{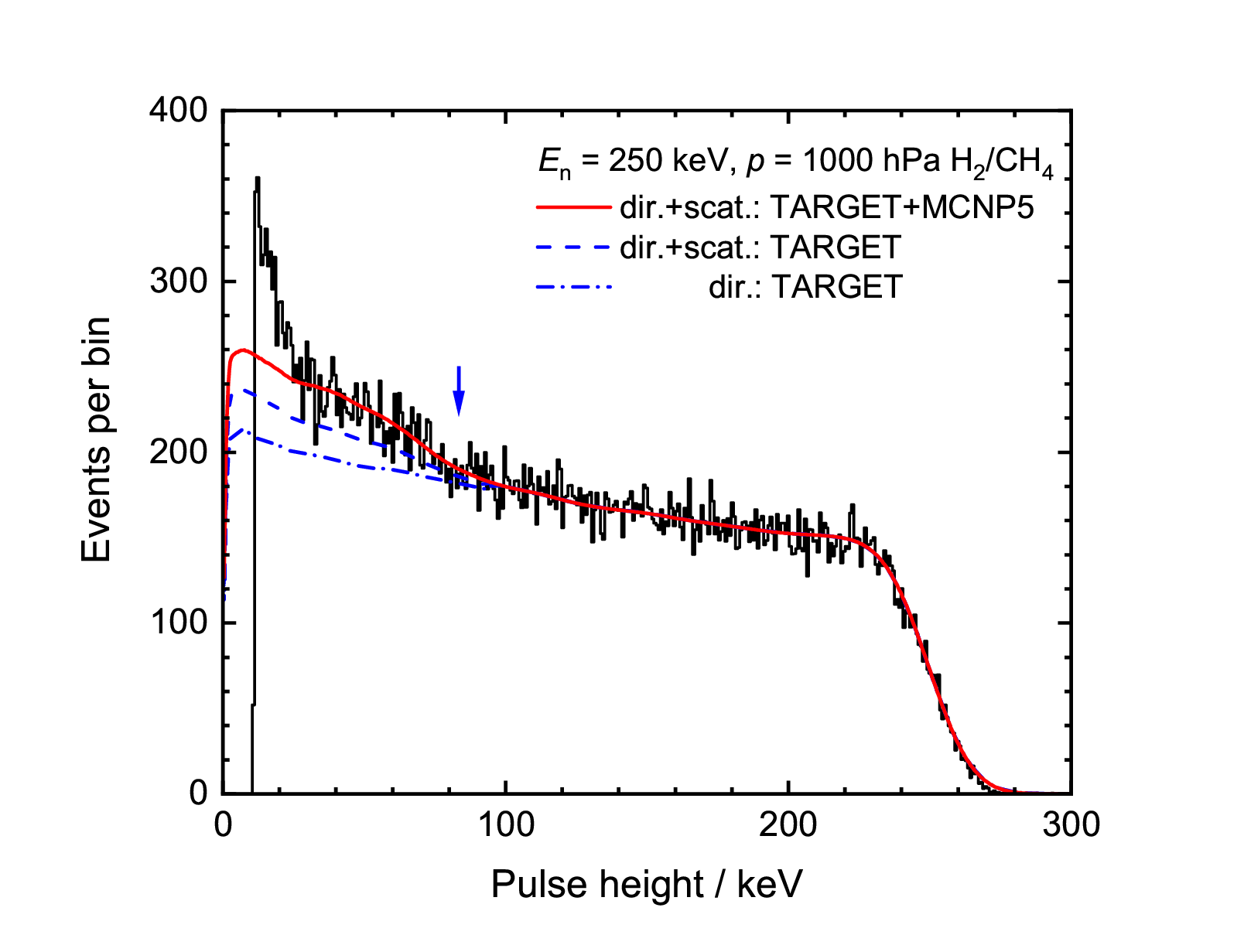}
  \centering
  \caption{\label{fig:scat-neutrons}
  Measured (black histogram) and calculated (lines) pulse height distributions calculated including the energy 
  distribution of neutrons scattered in the target. The simulations were carried out with MCNP5 (red solid line) 
  and TARGET (blue dashed line). The dashed-dotted line shows the contribution from the direct neutrons only.
  The blue arrow marks the lower bound of the fit region.}
\end{figure}

The effect of scattered neutrons on the pulse height distribution is exhibited in figure\,\ref{fig:scat-neutrons} for 250\,keV 
neutrons produced with a lithium target on a tantalum backing. As discussed in section \ref{sec:Fields}, the energy 
distribution calculated with MCNP5 improves the agreement of the measured and the calculated pulse height distribution 
in the region below 100\,keV.

\begin{figure}[ht]
  \includegraphics[width=0.45\textwidth]{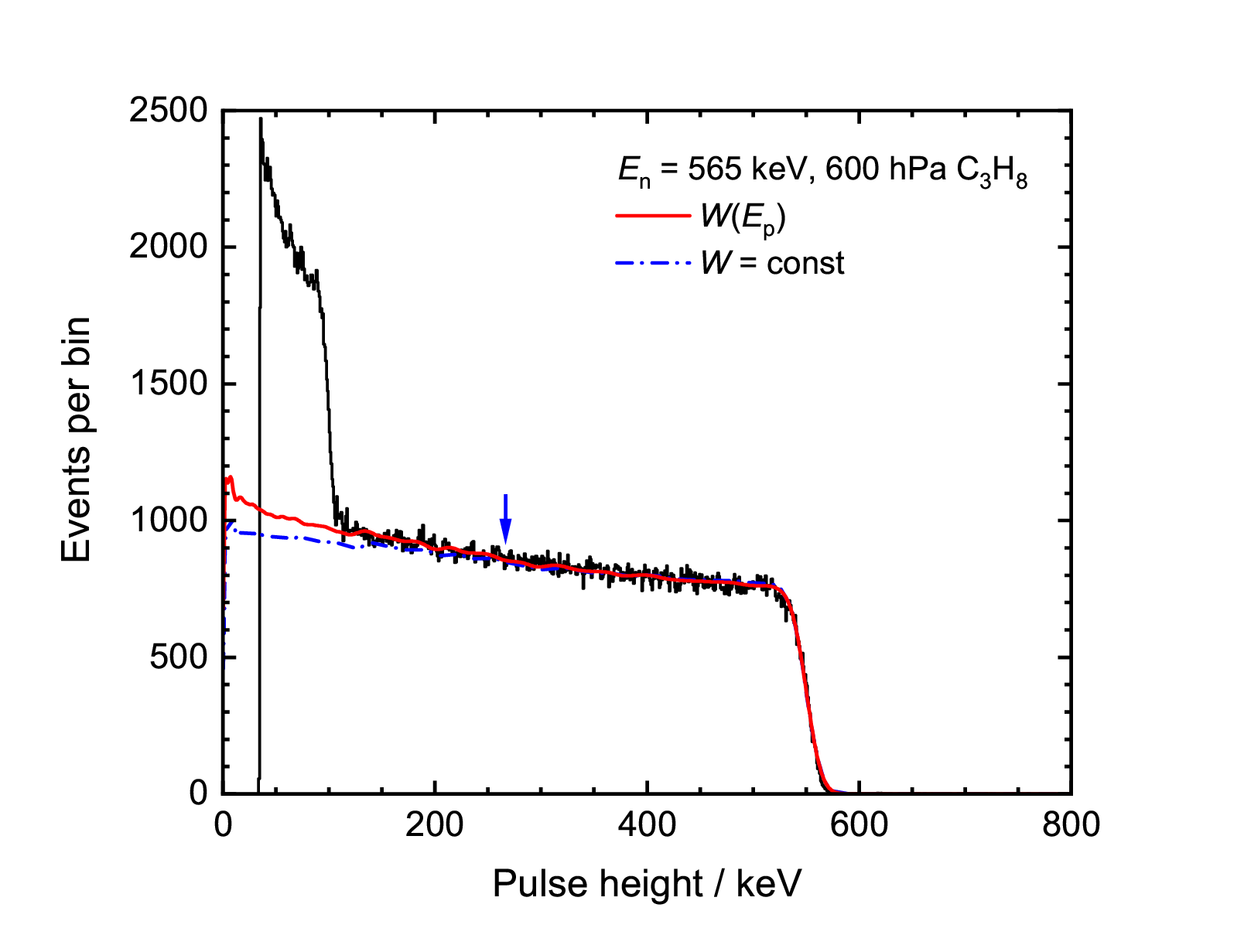}
  \centering
  \caption{\label{fig:W-value}
  Measured (black histogram) and calculated (lines) pulse-height distributions calculated using the 
  energy-dependent relative $W$-value data shown in figure\,\ref{fig:W-valuedata} (red solid line) and an 
  energy-independent $W$-value (blue dashed-dotted line). The  fit was restricted to the pulse height 
  region above 270\,keV. The edge at about 100\,keV is due to carbon recoil nuclei. Recoil nuclei 
  with $Z > 2$ are not transported by MCNPX. Hence, this structure is missing in the calculated 
  pulse height distribution.}
\end{figure}

Figure\,\ref{fig:W-value} shows the effect of an energy-dependent $W$-value for propane on the pulse-height 
distribution for 565\,keV neutrons. The slope of the experimental distribution is only reproduced when the 
energy-dependent \textit{W}-value is employed. This makes it possible to use the entire pulse-height region 
above the carbon recoil edge for the analysis and reduces the uncertainty due to the selection of the fit region.

\subsubsection{\label{subsubsec:RPPC-Analysis}Analysis of  P2 Measurements}

The neutron fluence measurements were analyzed by fitting a calculated pulse height distribution to 
the measured one after correcting for the contribution of neutrons scattered in air and on structural 
materials. The loss of events due to the dead time was also included in the corrections. For measurements with propane, 
the fit region was restricted to the pulse height region above the carbon recoil edge, corresponding 
to a proton energy of about $0.29\,E_\mathrm{n}$. With hydrogen-methane mixtures, the lower 
bound of the fit region was set to about $0.15\,E_\mathrm{n}$ to avoid interference from events produced  
by 478\,keV photons from \textsuperscript{7}Li(p,p'$\gamma$).  

The yield $Y_\mathrm{n,dir} = (\mathrm{d}N_\mathrm{n,dir}/\mathrm{d}\Omega)$ of direct neutrons emitted 
from the target is
\begin{equation}
  \label{eq:YieldP2}{
  Y_\mathrm{n,dir} = \frac {k_\tau \, N_\mathrm{p}} {k_V \, k_{pT} \, k_\mathrm{fit} \,  C_\mathrm{p}} d^2.
  }
\end{equation}
The net number of detected recoil proton events in the fitting region after subtraction of the events registered
with the shadow cone in place is $N_\mathrm{p}$ and 
$C_\mathrm{p}$ denotes the number of simulated events in the same region per unit fluence $\Phi_\mathrm{n,dir}$ 
of simulated direct neutrons in vacuum at the nominal distance $d_0 \approx d$ from the source.
The correction factors $k_\tau$ and $k_V$ account for dead time losses and  for the deviation of the effective 
sensitive volume from the geometrical sensitive volume $V_\mathrm{geo}$ and $k_{pT} = (p / p_0) (T_0 / T)$ denotes 
the correction for the deviation of the temperature and pressure of the counting gas from the reference 
values $p_0$ and $T_0$ used for the simulation. The correction factor $k_\mathrm{fit} = 1$ is used to 
introduce the uncertainty resulting from the selection of the fit region. The actual distance of the center 
of the sensitive volume from the target is $d$.

To evaluate the uncertainty of the Monte Carlo simulation, $C_\mathrm{p}$ can be broken down into the contribution 
from an ideal point-like detector at distance $d_0$, with volume $V_\mathrm{geo}$ and hydrogen density $n_\mathrm{H}$, 
multiplied by a kernel factor $G_\mathrm{MC}$ that describes the deviation of the Monte Carlo simulation from the result for an 
ideal detector,
\begin{equation}
  \label{eq:MCP2}{
  C_\mathrm{p} =  k_E \, k_\mathrm{sc,f} \, G_\mathrm{MC} \, n_\mathrm{H} \, V_\mathrm{geo} \, \sigma_\mathrm{np}(E_\mathrm{n,dir}).
  }
\end{equation}
The n-p scattering cross section at the mean energy $E_\mathrm{n,dir}$ of the direct neutrons is 
denoted by $\sigma_\mathrm{np}$.  The hydrogen density $n_\mathrm{H}$ is calculated for the 
reference conditions $p_0$ and $T_0 = 293.15\,\mathrm{K}$ using an equation-of-state for real 
gases \cite{DIN_1980}. The correction factor $k_\mathrm{sc,f}$ accounts for the additional
simulated recoil proton events in the fit region produced by neutrons scattered in the target.
The correction factor $k_E = 1$ is used to introduce the uncertainty of the n-p cross section 
$\sigma_\mathrm{np}(E_\mathrm{n})$ 
due to the uncertainty of the mean energy $E_\mathrm{n,dir}$ of the direct monoenergetic neutrons. 
For the reasons discussed in \ref{subsec:Production}, this contribution is only relevant for 
measurements with P2 in neutron fields produced using the \textsuperscript{7}Li(p,n)\textsuperscript{7}Be reaction. 
For the fields produced with the \textsuperscript{3}H(p,n)\textsuperscript{3}He  reaction, 
the relative uncertainty of $\sigma_\mathrm{np}$ is smaller than $2\times10^{-3}$.

The uncertainty of $G_\mathrm{MC}$ was assessed by voiding all elements of the model except for the counting 
gas and those elements for which the effect on the number of recoil proton events was investigated, for example the entrance 
window, the cylindrical housing, the cathode or the anode support. These numbers were compared with those for a model 
that was fully voided except for the counting gas. The relative difference was multiplied by the relative uncertainty 
of the total or differential cross section for the respective material to obtain the uncertainty contribution to 
$G_\mathrm{MC}$, which was added quadratically to those for the other elements of the model. 
Table\,\ref{tab:Unc.P2} shows the relative uncertainties for the quantities appearing in (\ref{eq:YieldP2}) 
and (\ref{eq:MCP2}).
\begin{table}
\caption{\label{tab:Unc.P2} Uncertainty contributions to the yield of direct neutrons measured with P2. 
The uncertainty contributions marked `Stat.' are related to variations from measurement to measurement 
while the contributions marked `Non-stat.' are constant or are strongly correlated over neutron energies.} 
\begin{indented}
\item[]
\lineup
\begin{tabular}{@{}llllll}
\br
Quantity             & Rel. uncertainty    & Stat.    & Non-stat.\\
\mr
$N_\mathrm{p}$       &  0.003\0 - 0.010    & $\times$ &          \\
$k_\tau$             &  0.0002             & $\times$ &          \\
$k_V$                &  0.0074             &          & $\times$ \\
$k_{pT}$             &  0.0020             & $\times$ &          \\
$k_\mathrm{fit}$     &  0.003\0 - 0.015    & $\times$ &          \\
$d$                  &  0.002\0            & $\times$ &          \\
$k_E$                &  0.004\0 - 0.008 \textsuperscript{a} & $\times$ &          \\
$k_\mathrm{sc,f}$    &  0.004\0 - 0.014    &          & $\times$ \\
$G_\mathrm{MC}$      &  0.005\0 - 0.013    &          & $\times$ \\
$n_\mathrm{H}$       &  0.002\0            &          & $\times$ \\
$V_\mathrm{geo}$     &  0.002\0            &          & $\times$ \\
$\sigma_\mathrm{np}$ &  0.005\0 - 0.009    &          & $\times$ \\
$C_\mathrm{p}$       &  0.015\0 - 0.023 \textsuperscript{b} &          & $\times$ \\
\mr                          
$Y_\mathrm{n,dir}$   &  0.015\0 - 0.028    &          &          \\   
\br
\end{tabular}
\item[] \textsuperscript{a} only for neutron fields produced with \textsuperscript{7}Li(p,n)\textsuperscript{7}Be, 
\item[] \textsuperscript{b} cumulated contributions from $k_E$, $k_\mathrm{sc,f}$, $G_\mathrm{MC}$, $n_\mathrm{H}$, 
$V_\mathrm{geo}$ and $\sigma_\mathrm{np}$
\end{indented}
\end{table}

\subsection{\label{subsec:RPT}Recoil Proton Telescope T1}

\subsubsection{\label{subsubsec:RPT-Design}Design of the Instrument}

\begin{figure}[ht]
  \includegraphics[width=0.45\textwidth]{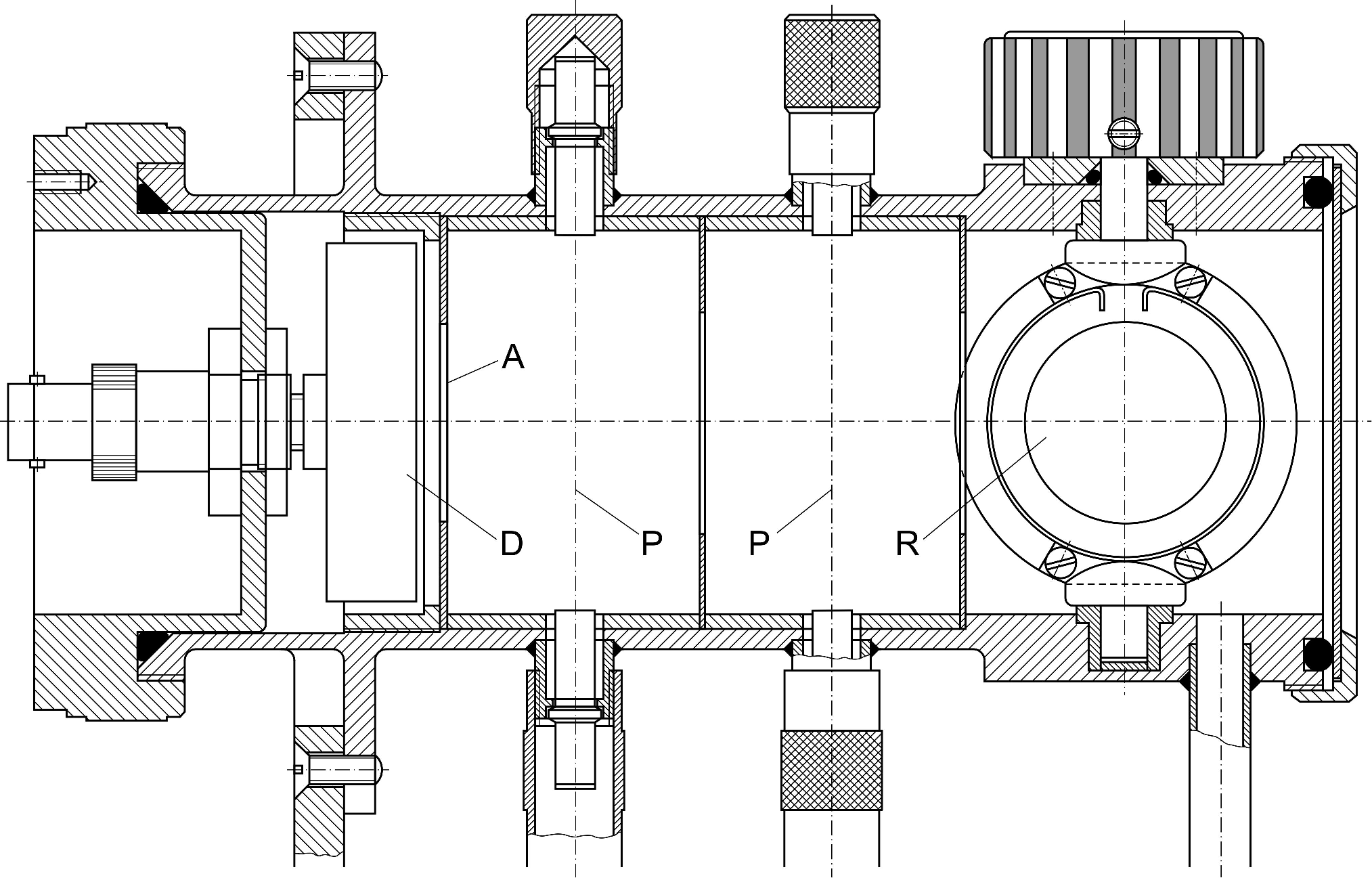}
  \centering
  \caption{\label{fig:T1}
  The recoil proton telescope T1 is used as the primary reference instrument for the neutron energy range 
  from 2\,MeV to 20\,MeV. The radiator with the tristearine layer is marked by R; the proportional counters and the silicon surface barrier
  detector are denoted by P and D, respectively. The tantalum aperture A in front of the silicon detector and 
  a mask directly in front of the tristearine layer define the solid angle for the detection of the recoil protons. 
  The radiator is shown in the intermediate \ang{90} position. For the foreground and background runs it is turned such that the 
  tristearine layer faces the detectors (\ang{0}) or the neutron source (\ang{180}), respectively. Depending on the 
  single event rates, T1 is used at two distances between the tristearine layer and the neutron source: 
  $d_1 = 22.61(2)\,\mathrm{cm}$ and $d_1 = 35.00(2)\,\mathrm{cm}$. The  distance $d_1$ is adjusted using calibrated gauges.}
\end{figure}

\begin{figure}[ht]
  \centering
  \begin{subfigure}[b]{0.45\textwidth}
    \includegraphics[width=\textwidth]{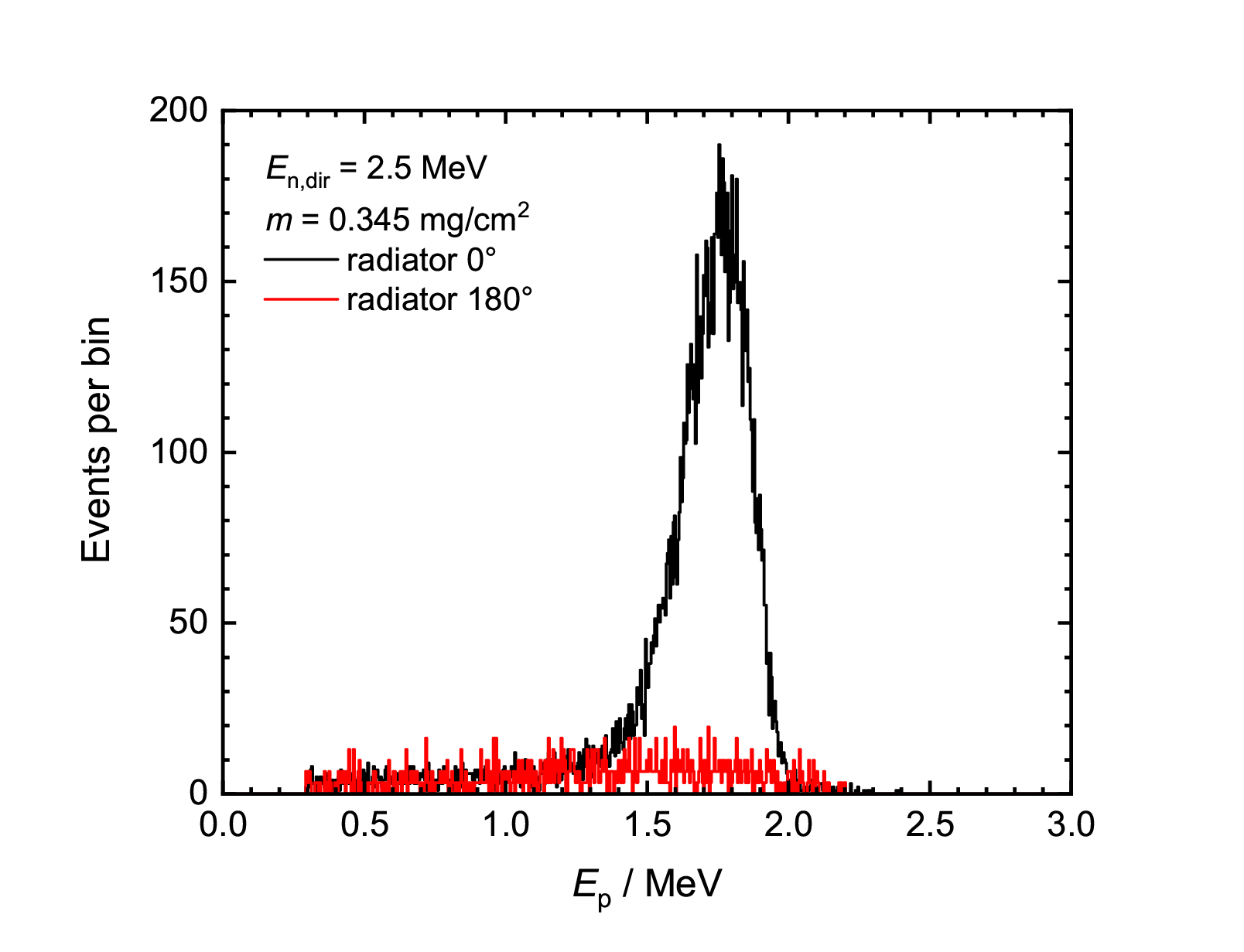}
  \end{subfigure}
  \begin{subfigure}[b]{0.45\textwidth}
    \includegraphics[width=\textwidth]{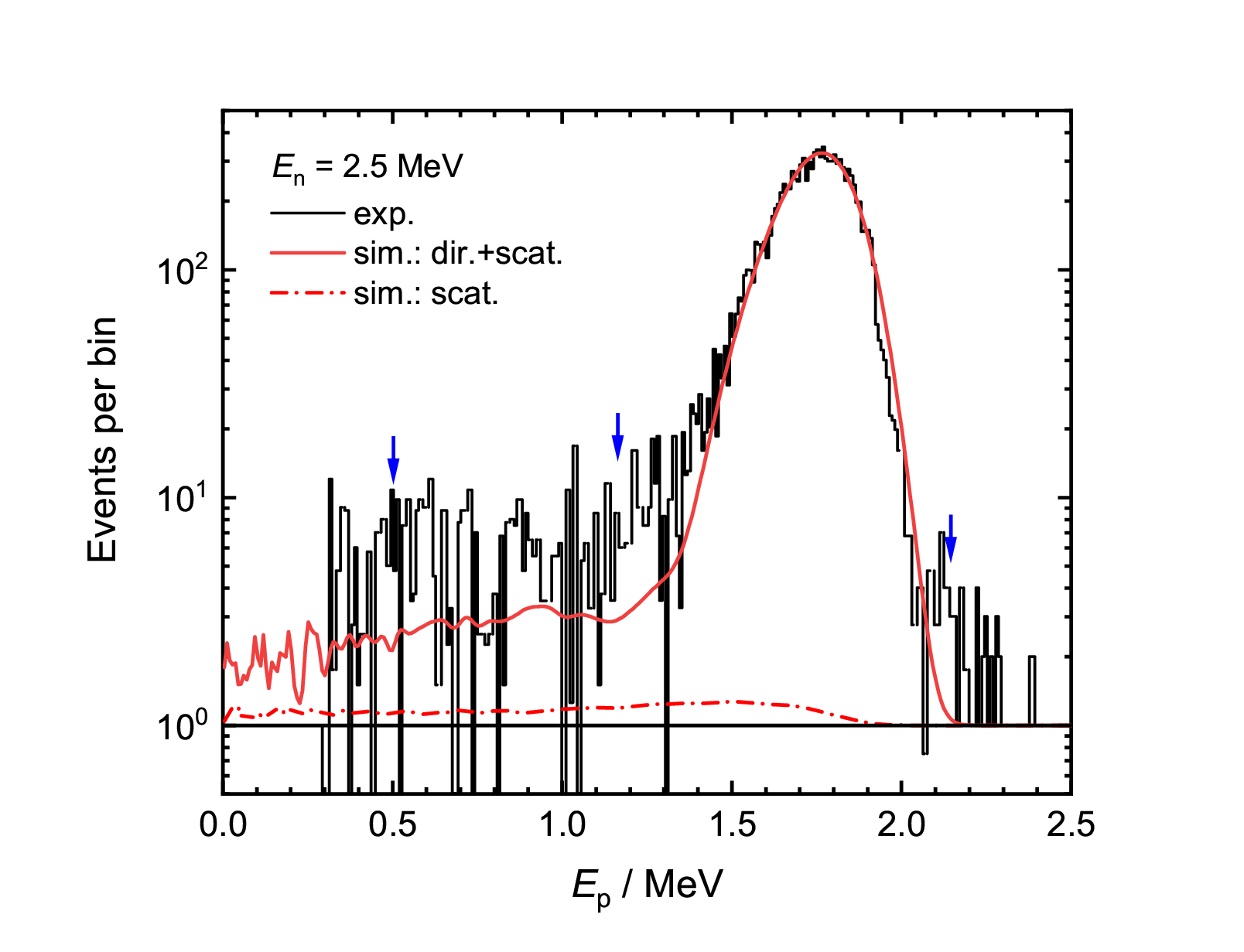}
  \end{subfigure}
  \caption{\label{fig:T1PHdistribution} 
  The upper panel shows foreground (black histogram) and background (red histogram) pulse height 
  distributions in the SB detector of T1 measured for 2.5\,MeV neutrons, using a 0.345\,mg/cm\textsuperscript{2} 
  tristearine radiator. In the lower panel, the experimental net pulse height distribution is 
  compared with the MCNPX simulation of T1. For better visibility of the tail region, two bins were grouped 
  together and the distributions were offset by one event per bin. The solid and dashed-dotted line show 
  all simulated events and those caused by neutrons scattered in the target assembly, respectively. 
  The blue arrows mark the peak and tail regions.
  The calculated number of events in the tail underestimates the measured number by 23\,\%.}
\end{figure}

The T1 RPT follows the design for RPTs for use in non-collimated (`open') neutron fields introduced by Bame \etal \cite{Bame_1957}.  
As shown in figure\,\ref{fig:T1}, the instrument consists of 
a hydrocarbon sample, the radiator, and a stack of two proportional counters (PC1 and PC2) using 
CO\textsubscript{2} as a counting gas together with a 1.5\,mm thick silicon surface barrier detector (SB). 
Depending on the neutron energy, 
the CO\textsubscript{2} pressure ranges between 60\,hPa and 300\,hPa. Neutrons incident on 
the radiator generate recoil protons that are detected in the detector stack. Valid events produce 
triple coincidences in PC1, PC2 and SB within a coincidence resolving time of about 2.5\,\unit{\micro s}. 
The solid angle for the detection of the recoil protons is defined by a tantalum aperture in front of the SB detector and
a mask directly in front of the tristearine layer that defines the effective diameter of  the radiator. 
In addition to valid recoil proton events originating from the radiator, neutron interactions with 
the counting gas and other parts of T1 produce single event rates in the three detectors which exceed the 
rate of triple-coincidence events by a factor of more than $10^3$. Accidental coincidences produced by these events are
measured in separate background runs with the radiator turned by \ang{180}, meaning towards the neutron source. 
The upper panel of figure\,\ref{fig:T1PHdistribution} shows foreground and background (radiator at 0° and 180°, respectively) 
pulse height distributions measured for 2.5\,MeV neutrons.

The radiators consist of tristearine 
($\mathrm{C}_{57}\mathrm{H}_{110}\mathrm{O}_{6}$)  layers produced by PVD
on 0.5\,mm tantalum backings. The masses per unit area of the layers were determined by differential 
weighing and measuring the aperture used to define the diameter of the layers during the evaporation. 
The stoichiometric composition was characterized by combustion analysis, $m_\mathrm{H}/m = 0.1235(10)$ by weight \cite{vanLeeuw_2015}. 

The width of the energy distribution of the valid recoil proton events 
depends on the thickness of the radiator. Therefore, a set of four radiators with masses per unit area ranging from 
0.756(6)\,mg/cm\textsuperscript{2} to 10.0855(28)\,mg/cm\textsuperscript{2} was used to cover 
the energy region 1.2\,MeV to 14.8\,MeV. As an example, the pulse height distribution measured with the SB detector 
for 2.5\,MeV neutrons is shown in figure\,\ref{fig:T1PHdistribution}.

\subsubsection{\label{subsubsec:RPT-Simulation}Monte Carlo Simulation}

The use of the RPT T1 with the D(d,n)\textsuperscript{3}He source and the PTB gas target has been reported in \cite{Siebert_1985}. 
The  dedicated Monte Carlo code SINENA \cite{Siebert_1985} developed specifically for measurements with this source was replaced by a 
full simulation of the neutron and charged-particle transport using MCNPX, v2.7 \cite{MCNPX_2011} for all neutron fields.  
In this way, the effect of neutron scattering in the RPT housing could be included consistently.

As for the P2 RPPC, the measurements with the T1 RPT were analysed by fitting a calculated pulse height 
distribution to the measured net pulse height distribution. The MCNPX model used for the simulation includes all 
relevant details of the instrument. Detailed field maps for the proportional counters were not available. 
Therefore, the effective sensitive volumes of the counters were modelled as ellipsoidal cylinders oriented 
parallel to the counting wires. Their sizes were adjusted such that the experimental pulse height distributions 
of the two proportional counters were reproduced.

For measurements carried out with solid-state targets, the neutron sources were modelled in the same way as for 
the P2 measurements, using the energy distributions of direct and scattered neutrons calculated with the TARGET code \cite{Schlegel_2005}. 
Only for the measurements with the D(d,n)\textsuperscript{3}He reaction and the deuterium gas target were the variation 
of the source yield along the axis of the gas volume and the anisotropy of the angular distribution included 
in the source description. 

Forced neutron interactions \cite{X-5_2004} were induced in the thin tristearine layers to 
obtain statistical uncertainties of less than 0.5\,\% in the Monte Carlo simulations.
Protons and alpha particles were transported with energy cutoffs of 1\,keV and 10\,keV, respectively. 
This makes sure that also events from \textsuperscript{16}O(n,$\alpha$)\textsuperscript{13}C reactions in 
the counting gas are properly tracked. The triple-coincidence events were reconstructed from the PTRAC files 
event by event using a dedicated filter code.
The finite pulse-height resolution of the silicon detector was accounted for by folding the 
simulated pulse height distributions with Gaussians of constant relative widths.

As demonstrated in the lower panel of figure\,\ref{fig:T1PHdistribution} 
the simulation also describes the tail on the left of the recoil peaks. This tail contains events 
produced by neutrons scattered in the target setup, the radiator backing or in the T1 housing together with events resulting from protons 
that lost energy in the counting wires or the edges of the apertures. 
These events do not contribute to the determination of the yield $Y_\mathrm{n,dir}$ of direct neutrons.
In this particular example, the calculated number of events in the tail underestimates the measured number by 23\,\%. 
This excess background is probably due to accidental coincidences and is assumed to extend into the peak region. 
For the very intense T(d,n) and D(d,n) sources, it can can be close to 50\,\% of the tail events. 
It is corrected by extrapolating the excess events into the peak region, assuming a linear decrease to zero within the peak region.

\subsubsection{\label{subsubsec:RPT-Analysis}Analysis of T1 Measurements}

The yield $Y_\mathrm{n,dir}$ of direct neutrons is obtained from the net number $N_\mathrm{p}$ of events in the peak region 
of the net SB pulse height distribution and the number $C_\mathrm{p}$ of events per unit source yield of direct neutrons 
in the corresponding region of the simulated pulse height distribution,
\begin{equation}
  \label{eq:YieldT1}
  Y_\mathrm{n,dir} = \frac {k_\mathrm{tail} \, k_\tau \, N_\mathrm{p}} 
  {k_\mathrm{d} \, k_\mathrm{fit} \, k_\mathrm{rel} \, C_\mathrm{p}}.
\end{equation}
Here, the correction factor $k_\mathrm{d}$ accounts for the difference between the actual distance between the target and 
the radiator and that assumed for the simulation. Its uncertainty includes the variation of the distance $d_1$ from the 
radiator to the beam spot on the solid-state target and the uncertainty due to the bending of the molybdenum foil of the gas target.
The correction factor $k_\mathrm{fit} \approx 1$ is introduced to include 
any effect of the fit region on the results and  $k_\mathrm{tail} \approx 1$ denotes a correction factor for an excess 
of events in the tail region which is extrapolated into the peak region. The correction factor 
\begin{equation}
  \label{eq:rel.correction}
  k_\mathrm{rel} = \frac {(\mathrm{d}\sigma_\mathrm{np}/\mathrm{d}\Omega_\mathrm{p,rel})}
  {(\mathrm{d}\sigma_\mathrm{np}/\mathrm{d}\Omega_\mathrm{p,n-rel})}
  = \frac {\gamma_0^2 (1 + \mathrm{tan}^2(\Theta_\mathrm{p}))} {1 + \gamma_0^2 \mathrm{tan}^2(\Theta_\mathrm{p})}
\end{equation}
corrects the effect of the non-relativistic kinematic formulas used in MCNPX for the transformation of the differential 
scattering cross section in the center-of-mass system to the recoil proton emission cross section in the laboratory 
system \cite{Thomas_1980}. Here, 
\begin{equation}
  \label{eq:gamma0}
  \gamma_0 = \frac {m_\mathrm{n}+m_\mathrm{p}+E_\mathrm{n}} {(m^2_\mathrm{n}+m^2_\mathrm{p}+2m_\mathrm{p}(m_\mathrm{n}+E_\mathrm{n}))^{1/2}}
\end{equation}
denotes the Lorentz factor for n-p scattering.
The correction factor $k_\mathrm{rel}$ is evaluated at the mean proton emission angle of $\bar \Theta_\mathrm{p}$ 
in the laboratory system and ranges from 1.001 at 2\,MeV to 1.008 at 14.8\,MeV. 

In the same way as for the RPPC, the number $C_\mathrm{p}$ of simulated events per unit source yield is broken down into 
the number of events expected for an ideal mass-less RPT and a kernel
factor $G_\mathrm{MC}$ that describes the neutron and proton transport effects in the simulation of the RPT,
\begin{equation}
  \label{eq:MCT1}
  C_\mathrm{p} = k_E \, k_\mathrm{sc,p} \, G_\mathrm{MC} \, \epsilon_\mathrm{geo} \, n''_\mathrm{H} \, 
  \sigma_\mathrm{np}(E_\mathrm{n,dir}).
\end{equation}
Here, $n''_\mathrm{H}$ denotes the number of hydrogen nuclei per unit area of the radiator and $\epsilon_\mathrm{geo}$ 
the geometric neutron detection efficiency of the RPT. The correction factor $k_\mathrm{sc,p} \approx 1$ 
accounts for neutrons scattered in the target and producing recoil proton events in the peak region.
The correction factor $k_E = 1$ is used to introduce the uncertainty of $\sigma_\mathrm{np}$ due to the uncertainty 
of the mean energy of the direct neutrons. This correction is only relevant for the neutron fields produced with the 
D(d,n)\textsuperscript{3}He reaction.

As shown by Sloan \etal \cite{Sloan_1982}, 
$\epsilon_\mathrm{geo}$ can be calculated numerically by integrating over the surfaces of the radiator and the aperture 
in front of the surface barrier detector, using the differential proton emission cross section as a weight. 
Thus, the uncertainty of the angular distribution around \ang{180} enters into the uncertainty of 
$\epsilon_\mathrm{geo}$ together with the uncertaintes of the mechanical dimensions. In order to separate the contributions 
resulting from mechanical 
dimensions and from nuclear data, the contribution from the relative angular distribution was added quadratically to that 
from the integrated cross section $\sigma_\mathrm{np}$, taken from the ENDF/B-V evaluation. 

Table\,\ref{tab:T1Efficiency} shows a comparison of geometrical efficiencies calculated numerically and with MCNPX 
using identical geometric parameters. The Romberg integration scheme was used for the numerical calculation, and the MCNPX model
was simplified accordingly by voiding all volumes except for a $100\,\unit{\micro g/cm^{2}}$ hydrogen radiator. 
The good agreement between the $N_\mathrm{p}/Y_\mathrm{n}$ values confirms 
the validity of the Monte Carlo calculations, especially the accuracy of the tabular cross section data at scattering 
angles of \ang{180} in the center-of-mass system.
\begin{table}
  \caption{\label{tab:T1Efficiency}
  Ratio of the quantity $N_\mathrm{p} / Y_\mathrm{n}$ calculated 
  with the Monte Carlo code MCNPX and using numerical integration. The relativistic correction factor $k_\mathrm{rel}$ was 
  used to correct for the use of  non-relativistic kinematics in MCNPX. The distance between the source and the radiator 
  surface of the RPT is denoted by $d_1$. The uncertainty is the statistical uncertainty of the Monte Carlo calcuation.}
  \begin{indented}
    \item[]
    \lineup
    \begin{tabular}{@{}ccll}
    \br
    $E_\mathrm{n}$ (MeV) &  $d_1$ (cm) & $k_\mathrm{rel}$ &  
    $  k_\mathrm{rel} (N_\mathrm{p}/Y_\mathrm{n})_\mathrm{MC} / (N_\mathrm{p}/Y_\mathrm{n})_\mathrm{num} $ \\
    \mr
   \01.2 &          22.539 &      1.0006 &      0.9955(27) \\
   \02.5 &          22.539 &      1.0013 &      1.0006(27) \\
   \05.0 &          37.239 &      1.0027 &      0.9973(31) \\
   \08.0 &          37.239 &      1.0043 &      1.0001(31) \\
    14.8 &          37.239 &      1.0079 &      1.0034(31) \\
    17.0 &          22.539 &      1.0091 &      1.0018(26) \\
    19.0 &          22.539 &      1.0101 &      0.9964(26) \\
    \br
   \end{tabular}
  \end{indented}
\end{table}
Table\,\ref{tab:Unc.T1} shows the relative uncertainties for the quantities appearing in (\ref{eq:YieldT1}) 
and (\ref{eq:MCT1}). The uncertainty of $G_\mathrm{MC}$ was evaluated in same way as that for the RRPC.
\begin{table}
  \caption{\label{tab:Unc.T1}
  Uncertainty contributions to the yield $Y_\mathrm{n,dir}$ of direct\ neutrons measured with T1.}
  \begin{indented}
    \item[]
    \lineup
    \begin{tabular}{@{}llllll}
      \br
      Quantity                 & Rel. uncertainty           & Stat.    & Non-stat. \\
      \mr
      $N_\mathrm{p}$           &  0.01\0\0 - 0.02           & $\times$ &           \\
      $k_\tau$                 &  0.0005   - 0.005          & $\times$ &           \\
      $k_\mathrm{tail}$        &  0.005\0  - 0.008          & $\times$ &           \\
      $k_d$                    &  0.002\0  - 0.003          & $\times$ &           \\
      $k_\mathrm{fit}$         &  0.005\0  - 0.008          & $\times$ &           \\
      $k_E$                    &  0.001\0  - 0.006          & $\times$ &           \\ 
      $k_\mathrm{sc,p}$        &  0.000\0  - 0.003          &          &  $\times$ \\
      $G_\mathrm{MC}$          &  0.007\0                              &          & $\times$  \\
      $n''_\mathrm{H}$         &  0.008\0  - 0.014                     &          & $\times$  \\
      $\epsilon_\mathrm{geo}$  &  0.003\0          \textsuperscript{a} &          & $\times$  \\
      $\sigma_\mathrm{np}$     &  0.010\0  - 0.018 \textsuperscript{b} &          & $\times$  \\
      $C_\mathrm{p}$           &  0.016\0  - 0.022 \textsuperscript{c} &          & $\times$  \\
      \mr
      $Y_\mathrm{n,dir}$       &  0.019\0  - 0.031                     &           &          \\
      \br
   \end{tabular}
  \item[] \textsuperscript{a} only uncertainty of mechanical dimensions, 
  \item[] \textsuperscript{b} including the uncertainty of  the angular distribution at \ang{180}
  \item[] \textsuperscript{c} cumulated contributions from $k_E$, $k_\mathrm{sc,p}$, $G_\mathrm{MC}$, 
                              $\epsilon_\mathrm{geo}$, $n''_\mathrm{H}$ and $\sigma_\mathrm{np}$
  \end{indented}
\end{table}

\subsection{\label{subsec:PLC}De\,Pangher Precision Long Counter LC1}

\subsubsection{\label{subsubsec:RLC-Design}Design of the Instrument}

\begin{figure}[ht]
  \centering
  \includegraphics[width=0.45\textwidth]{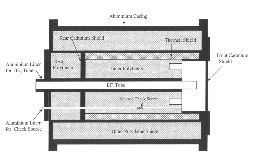}
  \caption{\label{fig:LC1} The De\,Pangher long counter LC1 of the PTB. Neutrons are incident from the right. The reference 
  surface for the measurement of the distance $d$ to the source is the front surface of the inner cylindrical 
  moderator layer surrounding the \textsuperscript{10}BF\textsubscript{3} proportional counter.}
\end{figure}

The De\,Pangher long counter \cite{DePangher_1966} LC1 of PTB is one out of a series of identical instruments manufactured
with very small tolerances \cite{Marshall_1970}. Other long counters of this series are used for neutron fluence measurements 
at several national metrology institutes. As shown in figure\,\ref{fig:LC1}, the long counter consists of a 
cylindrical polyethylene (PE) moderator surrounding a proportional counter filled with BF\textsubscript{3} enriched in \textsuperscript{10}B.
An intermediate cylindrical layer is made of borated PE and acts as a shield against external thermal neutrons. 

The long-term stability of LC1 is regularly checked with an Am-Be neutron source inserted into a channel 
in the inner PE moderator layer. After correction for the decay of \textsuperscript{241}Am, the event rates measured with 
this source showed a relative standard deviation of 0.0035 over the period from 2010 to 2022.

\subsubsection{\label{subsubsec:PLC-Response}Calculation of the Fluence Response}

The analysis of the measurements performed with LC1 is largely based on the extensive studies of long counters  
carried out at the National Physical Laboratory (NPL) in Teddington (UK) \cite{Roberts_2004}. Therefore, the MCNP model developed 
at the NPL was also adopted to model the LC1 PLC for the present study. A recent review of this model by N.J.\,Roberts 
(NPL) showed that the borated PE of the thermal shield was modelled assuming a boron content of 5\,\% of \textsuperscript{10}B 
per unit mass \cite{Roberts_2019}. This is not consistent with the available information \cite{Roberts_2019} which suggests 
a mass fraction of 5\,\% of \textsuperscript{nat}B per unit mass instead. 
Therefore, the model for LC1 used for this study was modified accordingly. In addition, a minor change was made for 
the PTB-specific mechanical support and the divergent source models also used for P2 and T1 were implemented.

\begin{figure}[ht]
  \centering
  \includegraphics[width=0.45\textwidth]{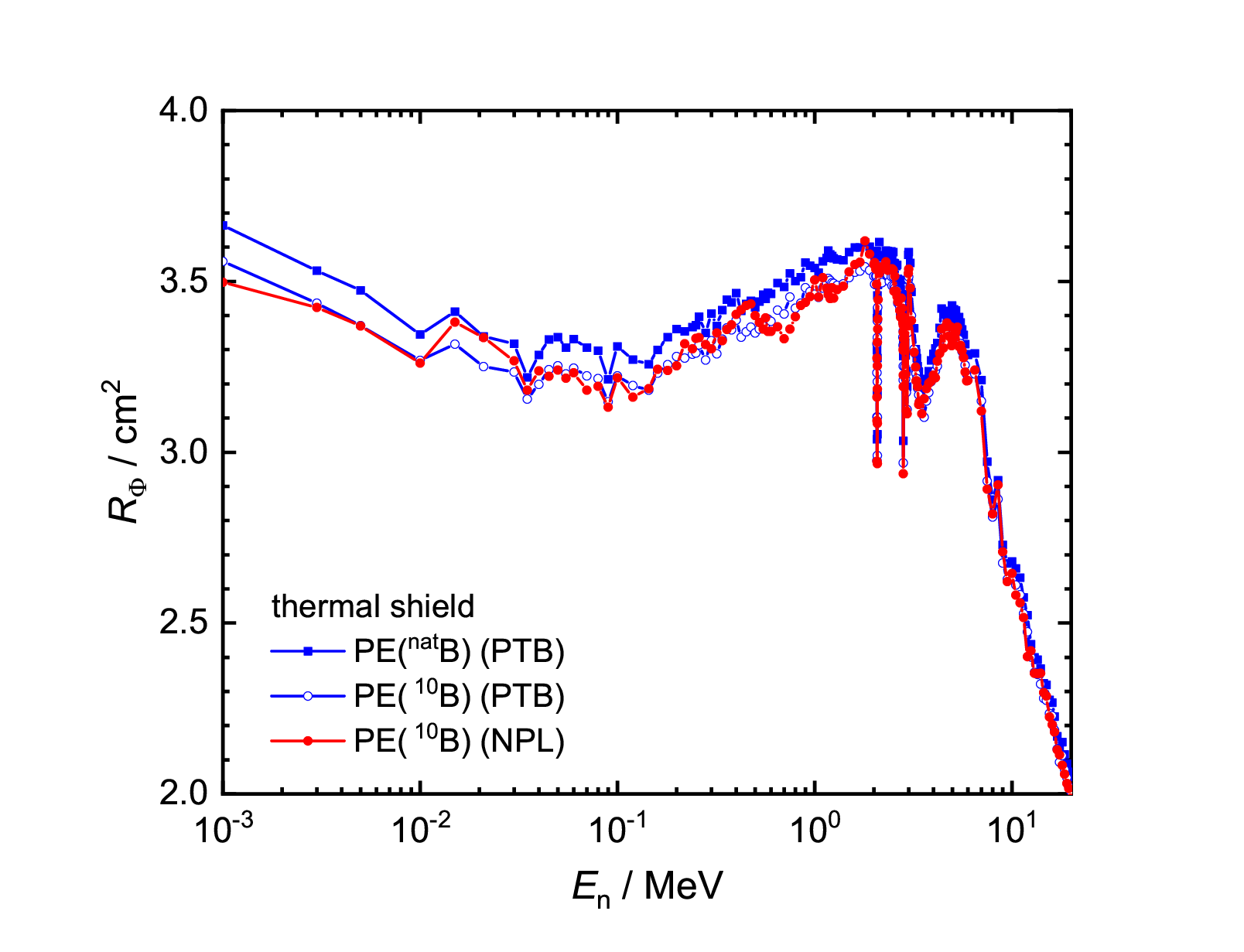}
  \caption{\label{fig:PLCresponse} Fluence response $R_\Phi$ of the PLC calculated with MCNP5 for two compositions of the borated PE 
  of the thermal shield. The data for 5\,\% \textsuperscript{10}B per unit mass (red closed circles) are the original 
  NPL data from \cite{Roberts_2004} 
  while the data for 5\,\% \textsuperscript{nat}B (blue closed squares) were calculated for the PTB's LC1 PLC. The open blue squares 
  show the results obtained for the same distribution of source neutrons but for borated PE with 5\,\% \textsuperscript{10}B. The 
  calculations were carried out for a unidirectional broad neutron beam, a BF\textsubscript{3} density of 
  $9.5414 \times 10^{-4}\,\mathrm{g/cm}^3$ and an \textsuperscript{10}B enrichment in the BF\textsubscript{3} of 96\,\%. 
  The  relative statistical uncertainties are smaller than 1\,\% (NPL data) and about 0.4\,\% (PTB data), respectively. 
  }
\end{figure}

Figure\,\ref{fig:PLCresponse} shows the fluence response $R_\Phi = N_\alpha / \Phi$ of the LC1 PLC, 
where $N_\alpha$ denotes the number of \textsuperscript{10}B(n,$\alpha_1+\alpha_2$) events. 
The fluence response data were calculated for a unidirectional 
extended neutron field and the two isotopic compositions of the borated PE 
in the thermal shield. All calculations were made for a nominal BF\textsubscript{3} 
density $\rho_\mathrm{nom} = 9.5414 \times 10^{-4}\,\mathrm{g/cm}^3$ and a \textsuperscript{10}B enrichment of 96\,\%. 
On average, the response calculated for the thermal shield containing \textsuperscript{nat}B is about 2.1\,\% higher 
than that for the original composition
assumed at the NPL. At energies below 1\,MeV, the deviation increases to 2.4\,\%. 

\subsubsection{\label{subsubsec:PLC-Calibration}Calibration}

\begin{figure}[ht]
  \centering
  \includegraphics[width=0.45\textwidth]{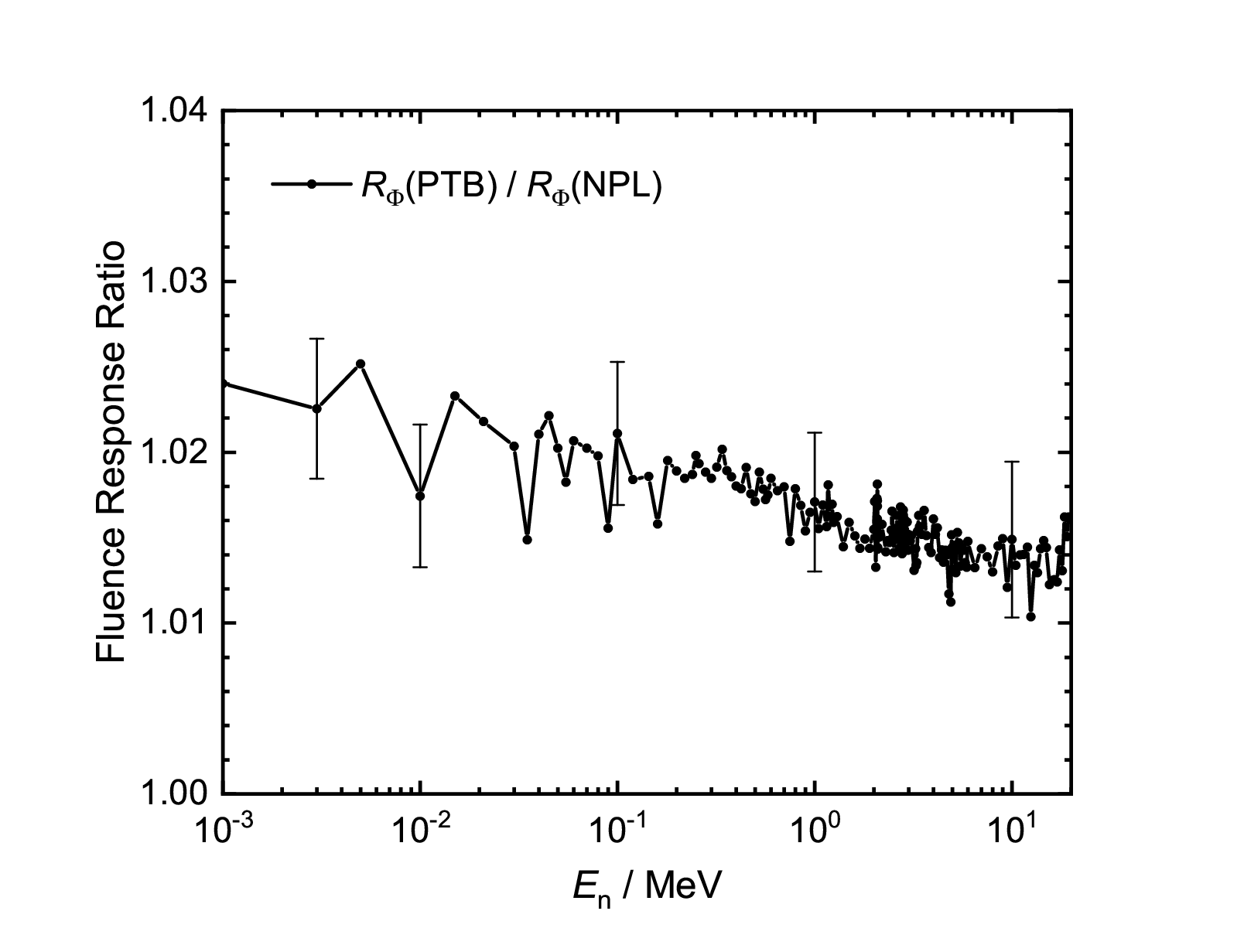}
  \caption{\label{fig:PLCresponseratio} Ratio of the fluence response $R_\Phi$ for the PTB's LC1 PLC (thermal shield: PE(\textsuperscript{nat}B), 
  BF\textsubscript{3} density:  $9.4818 \times 10^{-4}\,\mathrm{g/cm}^3$) to that for the NPL's PLC (thermal shield: PE(\textsuperscript{10}B), 
  BF\textsubscript{3} density:  $9.5414 \times 10^{-4}\,\mathrm{g/cm}^3$). The data for the NPL's PLC are shown by the 
  red solid circles in figure \ref{fig:PLCresponse}.
  In order  to reduce the statistical uncertainty of the ratios, 
  the two MCNP5 calculations were carried out for the same distribution of source neutrons. The error bars 
  are shown for representive energies only. They indicate the 
  relative statistical uncertainty of about 0.4\,\%.  
  }
\end{figure}

The fluence response of a long counter is calibrated by adjusting the nominal 
BF\textsubscript{3} density in the proportional counter such that measurements in 
known neutron fields are reproduced on average. At the NPL, a set of six radionuclide neutron sources 
was used. These sources had been calibrated in the NPL manganese bath \cite{Axton_1965}, \cite{Roberts_2010_1}. 
At PTB, only \textsuperscript{252}Cf 
sources are available for this purpose. The emission rates of these sources are traceable to the NPL manganese bath and 
the correction factors $k_\mathrm{an}(\Theta)$ for the angular anisotropy of the sources were measured at PTB. 
Measurements with two different sources were 
carried out in 2012 and 2023 in the PIAF low-scatter hall. For each source, two measurements were made at different distances. 
Table\,\ref{tab:Cf} shows the correction $k_\mathrm{Cf}$ for the response of LC1 in the \textsuperscript{252}Cf 
prompt fission neutron spectrum (PFNS),
\begin{equation}
    k_\mathrm{Cf} = 4\pi \frac {\dot Y_\mathrm{nom}} {k_\mathrm{an}(\ang{90})\,B},
\end{equation}
where $B$ denotes the emission rate of the \textsuperscript{252}Cf source and $\dot Y_\mathrm{nom}$ the source yield rate calculated using the nominal 
BF\textsubscript{3} density $\rho_\mathrm{nom}$. The generalized weighted mean \cite{Cox_2006} of the four 
measurements is $\bar{k}_\mathrm{Cf} = 0.995(8)$. It accounts for correlations induced by using 
the same \textsuperscript{252}Cf source for the measurements at the two distances, by the MCNPX simulation and by the long-term 
stability of LC1. To properly account for the dependence of the partial self-shielding in the BF\textsubscript{3} counter on the energy of the 
incident neutrons, the correction factor was not applied directly to the response calculated for the nominal BF\textsubscript{3} 
density $\rho_\mathrm{nom}$, but the BF\textsubscript{3} density was reduced to  $9.4818 \times 10^{-4}\,\mathrm{g/cm}^3$. 
This brings the correction factor to unity, $\bar k_\mathrm{Cf} = 1.000(8)$. Its uncertainty indicates the uncertainty of the normalization of the LC1 response 
to the \textsuperscript{252}Cf emission rate. Figure \ref{fig:PLCresponseratio} shows the ratio of the fluence response calculated for LC1 to that 
calculated for the NPL's PLC, that is, for a thermal shield made of  PE(\textsuperscript{10}B) and the nominal BF\textsubscript{3} density 
$\rho_\mathrm{nom} = 9.5414 \times 10^{-4}\,\mathrm{g/cm}^3$. In the energy range between 100\,keV and 20\,MeV, the residual energy dependence 
of the response ratio amounts to about 0.5\,\%.

\begin{table*}
  \caption{\label{tab:Cf} Correction factor $k_\mathrm{Cf}$ for the response of the LC1 long counter in the \textsuperscript{252}Cf PFNS. 
  The distance from the center of the \textsuperscript{252}Cf source to 
  the surface of the moderator surrounding the BF\textsubscript{3} counter is denoted by $d_0$. 
  The uncertainties indicated for $k_\mathrm{Cf}$ comprise the statistical uncertainty of the measured event rate $\dot N_\alpha$, 
  the effect of the field divergence and the total uncertainty of the emission rate $B$ of the \textsuperscript{252}Cf sources.}
  \begin{indented}
    \item[]
    \lineup
    \begin{tabular}{@{}llllll}
      \br
      Date       &   Source   & $d_0$ (cm)  &  $ B (10^5\,\mathrm{s}^{-1}$) & $ \dot N_\alpha (\mathrm{s}^{-1}$) & $k_\mathrm{Cf}$  \\
      \mr
      2012-03-12 &   3524NC   &     227.7   & \07.28(11)  & \03.713(15) & 0.990(17)  \\
      2012-03-14 &   3524NC   &     372.7   & \07.27(11)  & \01.363(8)  & 0.983(17)  \\
      2023-01-22 &   NS0520   &     240.8   &  38.68(27)  &  17.833(12) & 0.998(9)   \\
      2023-01-29 &   NS0520   &     372.7   &  38.49(27)  & \07.295(9)  & 0.994(9)   \\
      \br
    \end{tabular}
  \end{indented}
\end{table*}

\subsubsection{\label{subsubsec:LC1-Analysis}Analysis of LC1 Measurements}

As for T1 and P2, the measurements with LC1 aim at determining the source yield $Y_\mathrm{n} = (\mathrm{d}N_\mathrm{n}/\mathrm{d}\Omega)$ 
and not the fluence $\Phi$ at a particular point in space. Therefore, it is not necessary to 
use the concept of the effective center for the analysis of measurements at arbitrary distances. Instead, the response 
of LC1 is calculated with MCNP5 for two standard distances using a divergent source model and including a cone filled with air
between the source and the long counter. The two standard distances $d_0$ between the reference surface of LC1 and the target are 
379.3\,cm and 247.3\,cm for neutron energies above and below 500\,keV, respectively.  For these distances, dedicated shadow cones 
are available for subtracting the events induced by neutrons scattered in air or from structural materials. 

The yield of direct neutrons $Y_\mathrm{n,dir}$ is calculated from the net number 
$N_\alpha$ of detected \textsuperscript{10}B(n,$\alpha$\textsubscript{0}+$\alpha$\textsubscript{1})\textsuperscript{7}Li 
events and the number $C_\alpha$ of simulated events per unit yield $Y_\mathrm{n}^\mathrm{sim}$ of source neutrons,
\begin{equation}
  \label{eq:YieldLC1}
  Y_\mathrm{n,dir} = \frac {k_\tau \, N_\alpha} {k_\mathrm{sc}\, \bar{k}_\mathrm{Cf} \, C_\alpha}.
\end{equation}
The correction factor $k_\mathrm{sc} = 1 + Y_\mathrm{n,sc}^\mathrm{sim} / Y_\mathrm{n,dir}^\mathrm{sim}$ 
accounts for the simulated events induced by scattered neutrons.

Table\,\ref{tab:Unc.LC1} shows the relative uncertainties for the quantities appearing in (\ref{eq:YieldLC1}).
The relative uncertainty of the fluence response $R_\Phi$ was 
investigated in detail by Roberts \etal \cite{Roberts_2004}. The first result of 0.014 \cite{Roberts_2004} was later increased to 
0.020 \cite{Gressier_2014} and this value is adopted for relative uncertainty of the quantity $C_\alpha$. 
\begin{table}
\caption{\label{tab:Unc.LC1} Uncertainty contributions to the yield of direct neutrons measured with LC1.} 
\begin{indented}
\item[]
\lineup
\begin{tabular}{@{}llllll}
\br
Quantity             & Rel. uncertainty    & Stat.    & Non-stat.\\
\mr
$N_\alpha$           &  0.001\0 - 0.003    & $\times$ &          \\
$k_\tau$             &  0.001\0            & $\times$ &          \\
$\bar k_\mathrm{Cf}$ &  0.008\0            & $\times$ &          \\
$k_\mathrm{sc}$      &  0.005\0 - 0.024    &          & $\times$ \\
$C_\alpha$           &  0.020\0 \textsuperscript{a}   &          & $\times$ \\
\mr                          
$Y_\mathrm{n,dir}$   &  0.022\0 - 0.032    &          &          \\   
\br
\end{tabular}
\item[] \textsuperscript{a} relative uncertainty adopted from the NPL contribution to \cite{Gressier_2014}
\end{indented}
\end{table}
The uncertainty of the correction factors for the effect of scattered neutrons are fully correlated for P2 
and LC1, meaning that the covariance $\mathrm{cov}(k_\mathrm{sc,f},k_\mathrm{sc})$ must be subtracted from the total variances 
when the uncertainty of the ratio of normalized yields $(Y_\mathrm{P2}/M_\mathrm{P2})/(Y_\mathrm{LC1}/M_\mathrm{LC1})$ is 
calculated. Here, $M_\mathrm{P2}$ and $M_\mathrm{LC1}$ denote the corrected numbers of monitor events for the respective measurements.

\section{\label{sec:Results}Results and Discussion}

\begin{figure}[htp]
  \centering
  \begin{subfigure}[b]{0.45\textwidth}
    \includegraphics[width=\textwidth]{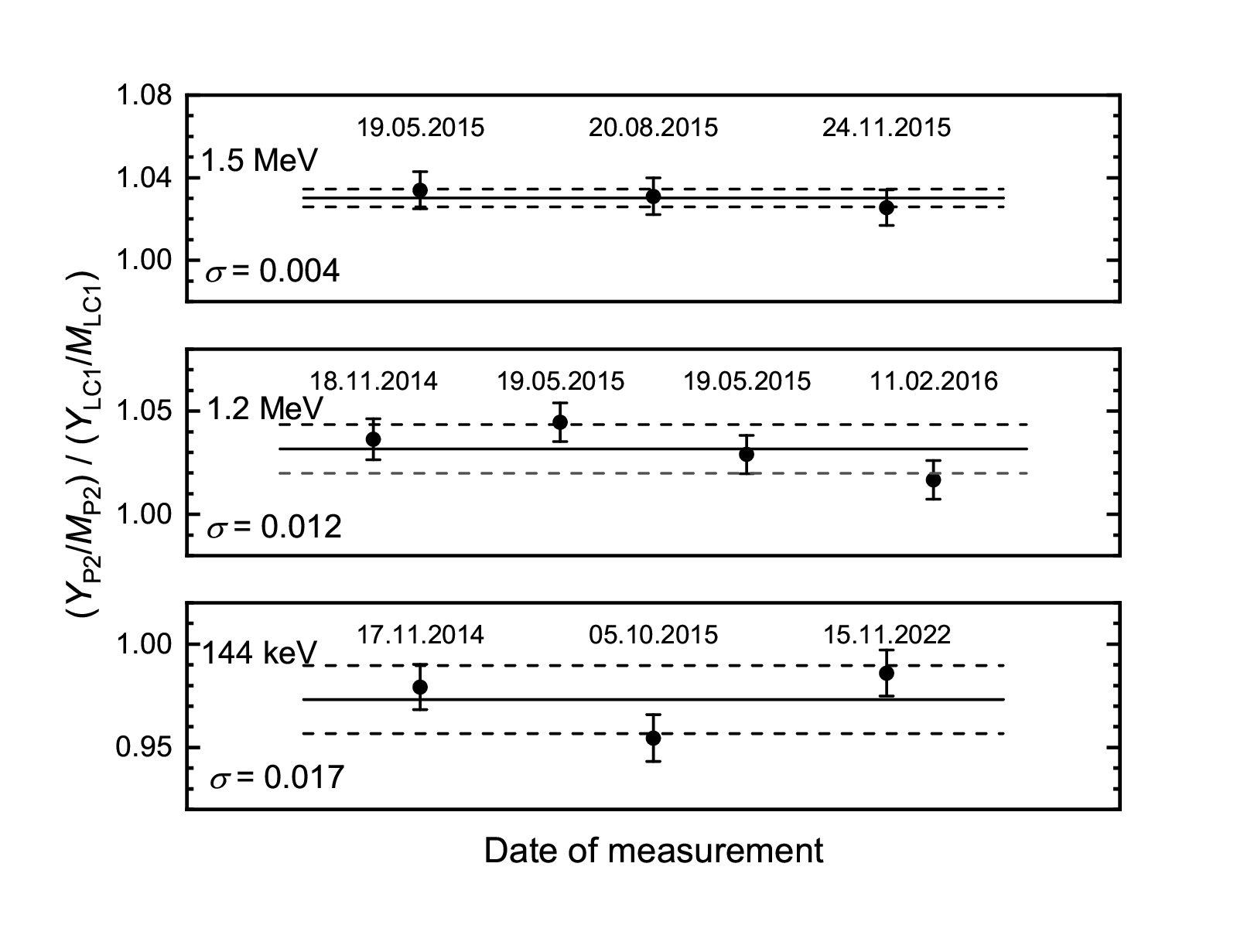}
  \end{subfigure}

  \begin{subfigure}[b]{0.45\textwidth}
    \includegraphics[width=\textwidth]{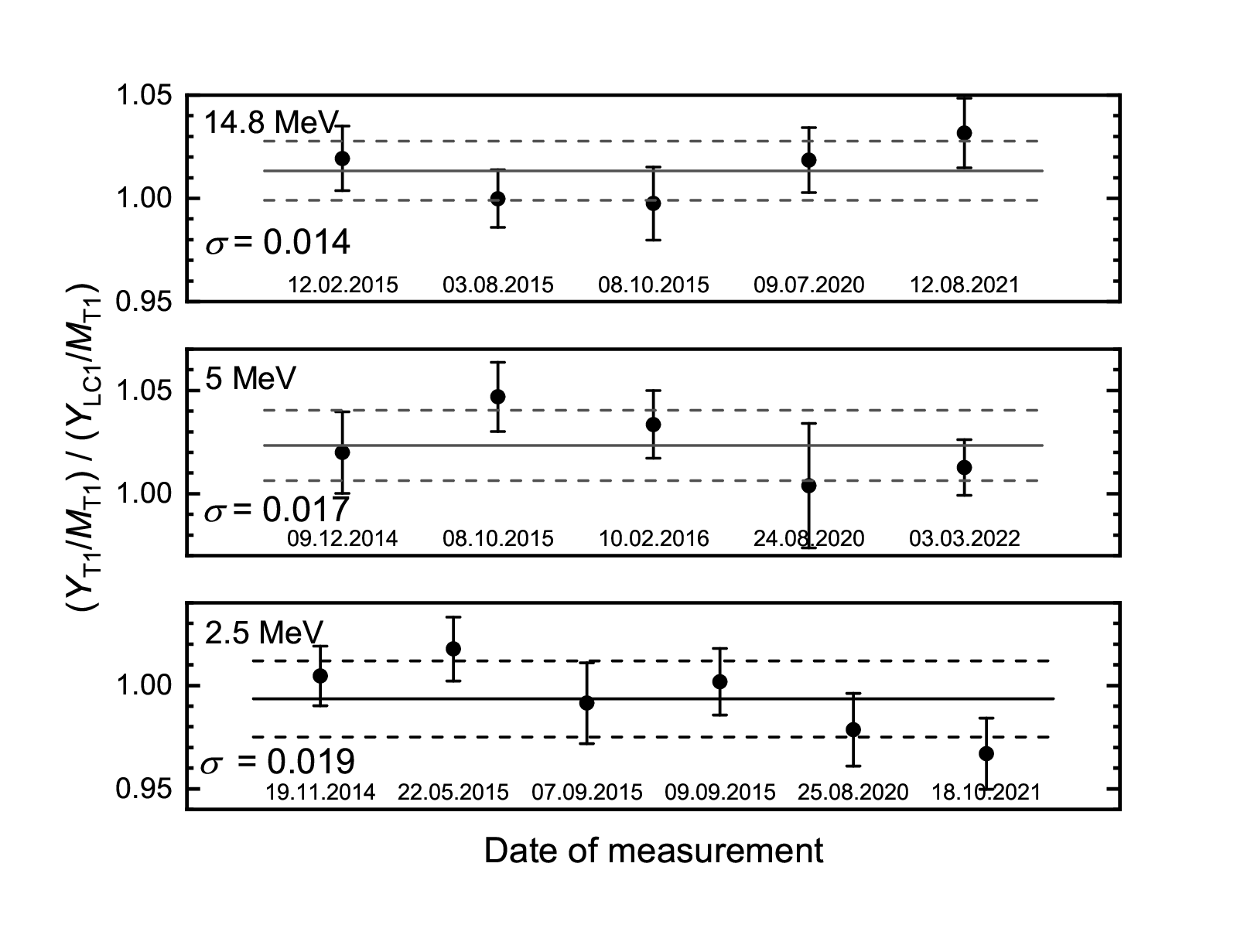}
  \end{subfigure}
  \centering
  \caption{\label{fig:VarianceP2T1} 
  Ratio of the target yields $(Y/M)$ of direct neutrons per event in the NM monitor measured with P2 (upper panel) and T1 (lower panel) 
  to those measured with LC1 as a function of the date of the measurement. The error bars show the statistical 
  uncertainty contribution. The solid and dashed lines exhibit the mean and the standard deviation of the 
  data points, respectively. For P2, the empirical standard deviation $\sigma$ of the data points ranges between 0.004 
  and 0.017 and for T1 between 0.014 and 0.019.}
\end{figure}

For selected energies, the reproducibility of the measurements for the period from 2016 to 2022 is demonstrated in 
figure\,\ref{fig:VarianceP2T1}. The error bars shown here comprise only those uncertainty components which are expected to
vary from measurement to measurement and are not correlated over neutron energy. The empirical relative standard deviation $\sigma$ of 
the yield ratios ranges between 0.4\,\% and 1.9\,\% and is consistent with the error bars of the data points.

\begin{figure}[htp]
  \centering
  \begin{subfigure}[b]{0.45\textwidth}
    \includegraphics[width=\textwidth]{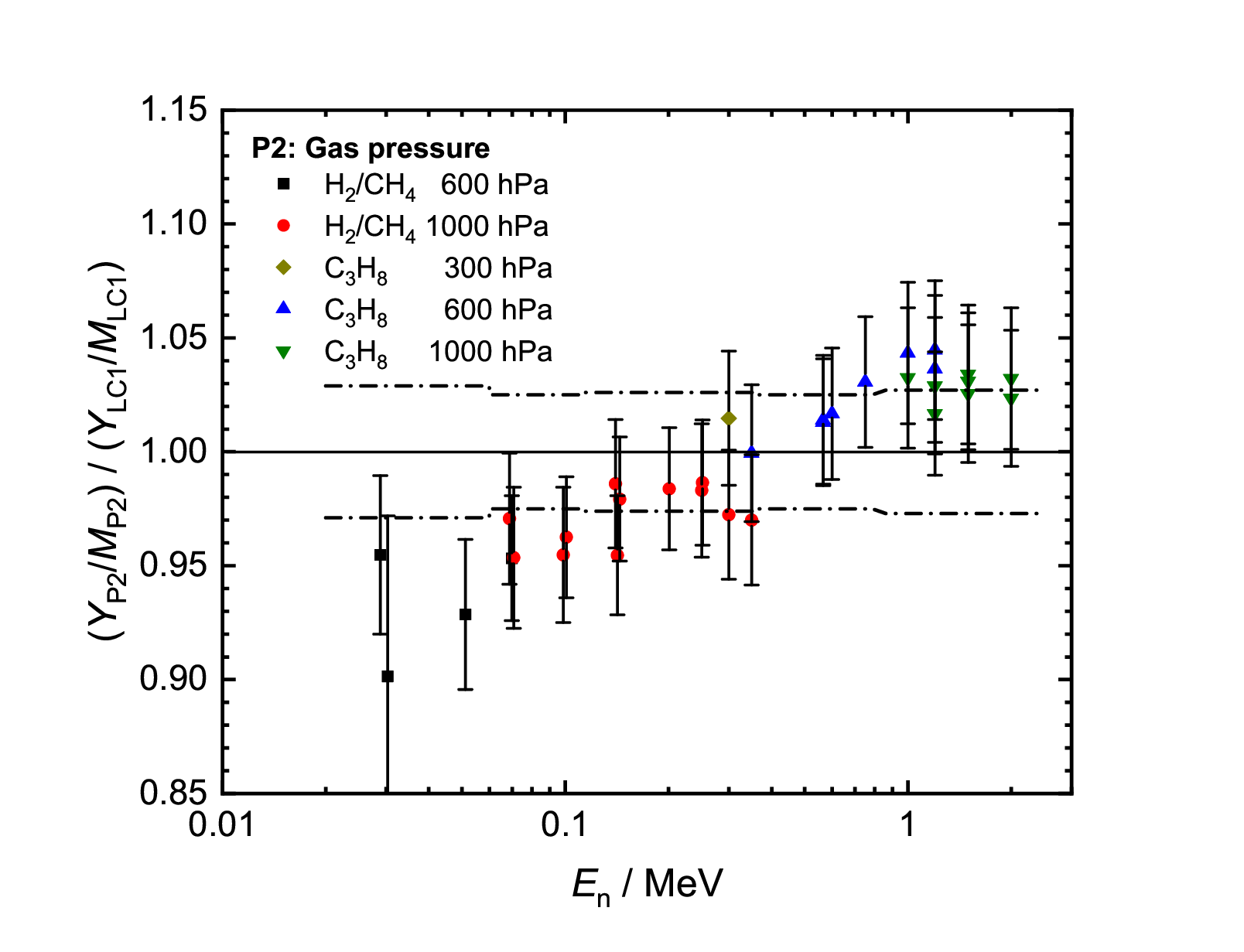}
  \end{subfigure}

  \begin{subfigure}[b]{0.45\textwidth}
    \includegraphics[width=\textwidth]{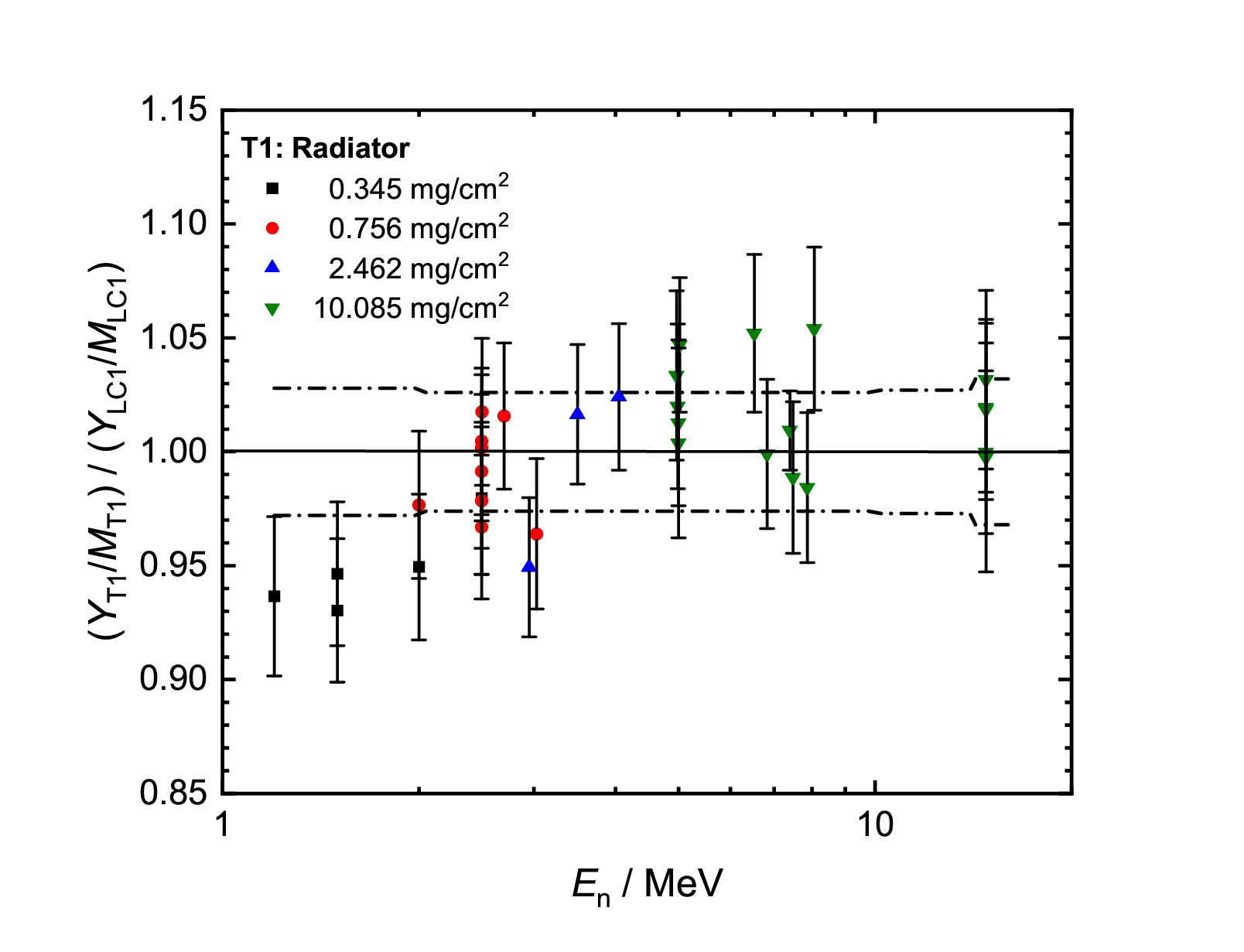}
  \end{subfigure}
  \centering
  \caption{\label{fig:RatiosP2T1} 
  Ratio of the target yields $(Y/M)$ of direct neutrons per event in the NM monitor measured with P2 (upper panel) and T1 (lower panel) to 
  those measured using LC1 for the same neutron field. The error bars show the total uncertainties, including the non-statistical 
  components which are correlated over neutron energies. The dashed-dotted lines show the non-statistical component of the 
  uncertainty of the measured yield ratios averaged over larger energy bins.}
\end{figure}

Figure\,\ref{fig:RatiosP2T1} shows the ratio of the target yields $(Y/M)$ per event in the NM monitor measured 
with P2 and T1 to those measured with LC1 for the same neutron field. The neutron energy ranges from 30\,keV to 
14.8\,MeV. As discussed above, the neutron fields with energies below 144\,keV were produced at neutron emission
angles $\Theta_\mathrm{n} > \ang{0}$. All other measurements were made at \ang{0}. The maximum neutron energy covered 
with P2 was 2\,MeV and the minimum neutron energy for T1 was 1.2\,MeV. The error bars indicate the total uncertainties 
and the dashed-dotted lines show the cumulated non-statistical uncertainty contributions (see tables\,\ref{tab:Unc.P2} 
and \ref{tab:Unc.T1}) averaged over larger energy bins.

The data measured with P2 and T1 match around 2\,MeV. Below 2\,MeV, the T1 data show a decrease while the P2 
results are almost constant between 1\,MeV and 2\,MeV. This indicates a problem with the RPT measurements at very 
low recoil proton energies. It could be caused, for instance, by the effect of electromagnetic interference at the 
low signal amplitudes found at these energies. It is also interesting to note that the yield ratios measured with P2 using 
the H\textsubscript{2}/CH\textsubscript{4} mixture as counting gas seem to be systematically lower than those measured using 
C\textsubscript{3}H\textsubscript{8}. This tendency needs further investigation.

For neutron energies above 100\,keV most of the present data are located 
between the non-statistical uncertainty margins indicated by the dashed-dotted lines. There is nonetheless a clear trend with neutron 
energy which continues at lower energies where the data fall below the lower uncertainty margin. 
Thus, the estimated uncertainties appear to be realistic for the measurements at an emission 
angle $\Theta_\mathrm{n} = \ang{0}$ . At angles greater than \ang{0}, 
however, there must be unrecognized sources of uncertainty. Possible explanations could be a contribution of 
scattered neutrons greater than that calculated with MCNP5 or an effect related to the increased kinematical sensitivity of the 
neutron yield and neutron energy on the emission angle $\Theta_\mathrm{n}$ for angles much larger than \ang{0}. This would 
also explain the increased scatter of the data below 144\,keV.

\begin{figure}[htp]
    \centering
    \includegraphics[width=0.45\textwidth]{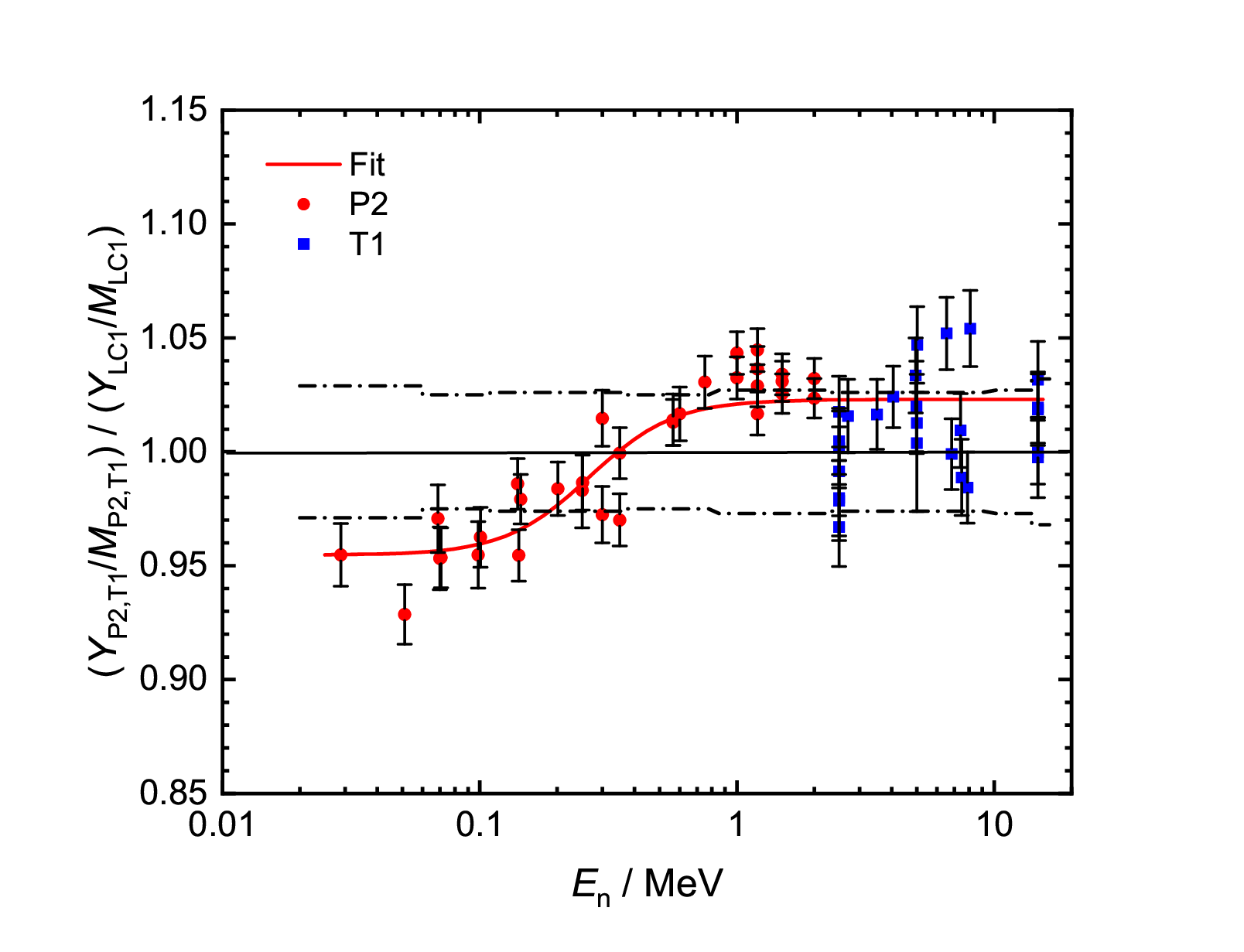}
    \caption{Ratio of the target yields $(Y/M)$ (red and blue closed symbols) fitted by a sigmoid function on a logarithmic energy scale (red solid line). 
    Data points measured with T1 at neutron energies below 2.5\,MeV and three other deviating data points at 
    30\,keV, 2.95\,MeV and 3.03\,MeV were excluded from the fit. The error bars show the statistical uncertainty component only. 
    The dashed-dotted lines exhibit the non-statistical component of the uncertainty of the measured yield ratios averaged over 
    larger energy bins.}
    \label{fig:FitP2T1}
\end{figure}

A reduced data set with error bars exhibiting the statistical uncertainty component only is shown in 
figure\,\ref{fig:FitP2T1}. The T1 measurements below 2\,MeV were discarded from this data set. In addition, 
it does not include the deviating P2 measurement at 30\,keV that was determined without photon subtraction. 
Nor does it include the T1 data points at 2.95\,MeV and 3.03\,MeV that are too close to the prominent dip in the 
\textsuperscript{12}C elastic scattering cross section around 2.98\,MeV which is also visible in the long counter 
response shown in figure\,\ref{fig:PLCresponse}.

The red solid line shows a fit of a sigmoid function on a logarithmic energy scale,
\begin{equation}
   \label{eq:log-sigmoid}
   S(E_\mathrm{n}) =
   \delta s \, \Bigl( \frac {2} {1 + \exp(-u / \sigma)} - 1 \Bigr) + s_0,
\end{equation}
with $u = \ln(E_\mathrm{n}/E_0)$, $E_0 = 0.27(5)\,\mathrm{MeV}$, $\sigma = 0.38(10)$, $\delta s = 0.034(4)$ 
and $s_0 = 0.988(4)$. 

Under the tentative assumption that the observed energy dependence of $(Y_\mathrm{P2,T1}/M_\mathrm{P2,T1})/(Y_\mathrm{LC1}/M_\mathrm{LC1})$ is due to 
the calculated energy dependence of the fluence response $R_\Phi$ of the long counter LC1, $S(E_\mathrm{n})$ describes 
the correction for the energy dependence of $R_\Phi(E_\mathrm{n})$ required to match the results obtained with P2 and T1. 

\begin{figure}[htp]
    \centering
    \includegraphics[width=0.45\textwidth]{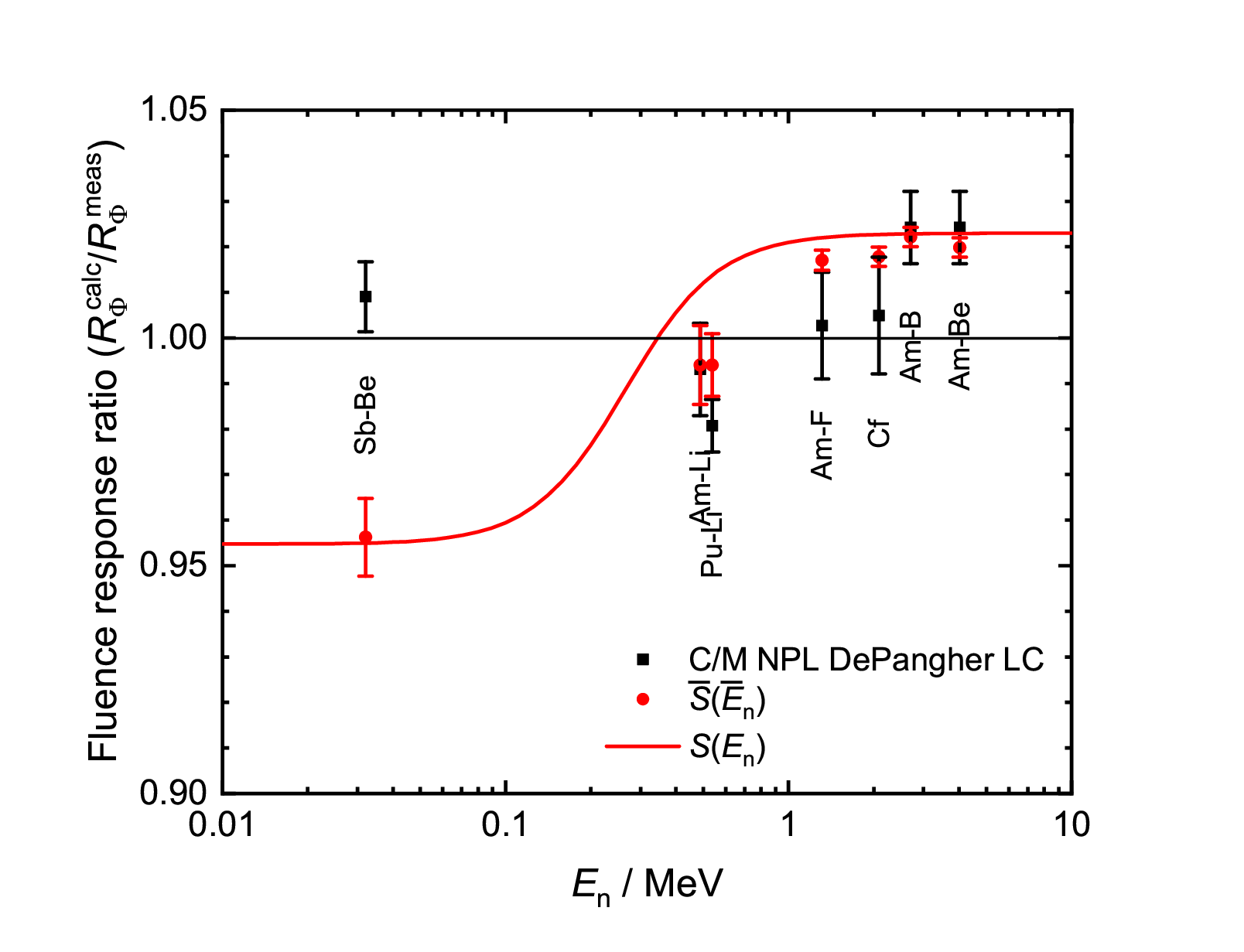}
    \caption{The black closed squares show the ratio of calculated and measured fluence response values for the NPL De\,Pangher 
    long counter and various radionuclide neutron sources \cite{Thomas_2023}. The red solid line is the fit $S(E_\mathrm{n})$ 
    to the ratio of normalized target yields $(Y/M)$ measured with P2 or T1 to that obtained with LC1. The red
    closed circles show the spectrum-averaged value $\bar S(\bar E_\mathrm{n})$ for the respective neutron sources. 
    The error bars of the spectrum-averaged data points were determined from the uncertainties of the fit parameters only.
    This means that they do not include the uncertainties of the energy distributions.}
    \label{fig:NPLsources}
\end{figure}

Therefore, it is interesting to compare $S(E_\mathrm{n})$ 
with the results of measurements carried out with the NPL De\,Pangher long counter and a set of radionuclide neutron 
sources which had been calibrated in the NPL manganese bath. Figure \,\ref{fig:NPLsources} shows a comparison of 
the ratio of calculated 
and measured (C/M) fluence response values for the various neutron sources in use at the NPL \cite{Thomas_2023} 
with the spectrum-averaged correction factor 
\begin{equation}
  \label{eq:averagedcorrectionfactor}
  \bar S(\bar E_\mathrm{n}) = \frac {\int\limits_{0\,\mathrm{MeV}}^{20\,\mathrm{MeV}}  S(E_\mathrm{n})\,R_\Phi(E_\mathrm{n})\,\Phi_E\,\mathrm{d}E_\mathrm{n}} 
  {\int\limits_{0\,\mathrm{MeV}}^{20\,\mathrm{MeV}}  R_\Phi(E_\mathrm{n})\,\Phi_E\,\mathrm{d}E_\mathrm{n}}.
\end{equation}
Here, $\Phi_E$ and $\bar E_\mathrm{n}$ denote the neutron energy distribution of the radionuclide source and its mean energy, respectively. 
This correction factor was calculated using the original 
MCNP model of the NPL, that is, with the thermal shield of PE(\textsuperscript{10}B) and a 
BF\textsubscript{3} density of $9.5414 \times 10^{-4}\,\mathrm{g/cm}^3$ adjusted by the NPL to give the best match to the measurements 
with the NPL radionuclide neutron sources.
It is interesting to note that the correction function $S(E_\mathrm{n})$ for the fluence response of the long 
counter seems to reproduce the results obtained at the NPL for the Am-Li, Am-F, Cf, Am-B and Am-Be sources. 
The results for the Pu-Li source and the Sb-Be photoneutron source, however, clearly deviate from this trend.

\section{\label{sec:Conclusions}Conclusions}

In the energy region between 144\,keV and 14.8\,MeV, this study confirms the consistency of neutron fluence measurements 
with the primary reference instruments P2 and T1 of the PTB and the LC1 PLC. 
The P2 RPPC and the T1 RPT can be used for neutron energies below and above about 2.2 MeV, respectively.
The observed energy dependence of the neutron yield ratios is within the estimated 
uncertainty margins and consistent with measurements with radionuclide neutron sources carried out at the NPL. This suggests 
that the deviation is actually caused by the long counter response calculated with MCNP.
The present results confirm the chain of traceability between fast neutron metrology based on the 
differential n-p scattering cross section as the primary standard and the manganese bath technique that relates 
neutron measurements to the activity standard. 
The fit to the measured yield ratios varies between 0.966 and 1.023, that is, around $0.995 \pm 0.029$, which is 
consistent with the uncertainty budgets of the instruments and 
the uncertainty of the calibration of LC1 with the PTB \textsuperscript{252}Cf neutron sources. 
This study is complementary to previous work \cite{Bardell_1994} using a \textsuperscript{252}Cf fission 
ionization chamber to compare neutron source emission rate measurements with the manganese bath and with 
a liquid scintillation detector operated in time-of-flight mode. 

At energies below 144\,keV, however, the results are less satisfactory. Here, the measured ratio of target yields is 
only about 0.948 and is clearly outside the calculated uncertainty margins. This deviation is currently not 
understood, but is most likely due to the energy distribution of the neutron fields produced by 
the \textsuperscript{7}Li(p,n)\textsuperscript{7}Be reaction 
at neutron emission angles greater than zero degrees. Further work is needed to characterize these fields 
in detail using time-of-flight spectrometry. Alternatively, neutron fields with energies between 8\,keV and 
about 40\,keV can be produced  using the 
\textsuperscript{45}\uppercase{S}c(p,n)\textsuperscript{45}\uppercase{T}i 
reaction \cite{Cosack_1987} at \ang{0}. It should, however, be noted that this only useful for metrological investigations 
and is no solution for routine 
applications because of the low neutron yield and the intense production of parasitic photons. Until the 
problem at neutron energies below 144\,keV is solved, the uncertainty of neutron yield measurements using 
P2 or LC1 at PIAF must be increased accordingly for these neutron fields. 

The present study demonstrates that fast neutron fluence measurements with relative uncertainties around 2\,\% are 
possible but challenging. This is more than sufficient for applications in radiation protection metrology but 
does not meet the most challenging demands from nuclear technology, especially for standard cross sections.

\ack
This study has a long lasting history dating back to the first draft of a PTB-Report from the early 1990s. 
Therefore, the authors are indebted to the former and present heads of the PTB neutron departments H.\,Klein, 
H.\,Schuhmacher and A.\,Zimbal for many discussions on this subject. The authors are grateful to N.J.\,Roberts and D.J.\,Thomas of 
the NPL for many valuable comments and for providing the MCNP model of the De\,Pangher long counter 
and the data for the NPL radionuclide neutron sources.

S.\,Löb, M.\,Thiemig and B.\,Schulze carried out the numerous measurements reported in this study and developed 
the preparation of the lithium targets. The PIAF accelerator staff provided the 
ion beams for the experiments. Their tireless effort is gratefully acknowledged.

\section*{References}
\bibliography{MET-102500.bib}

\end{document}